\documentclass[aps]{quantumarticle}

\usepackage{graphics}
\usepackage{epsfig}

\usepackage{amsmath}

\def\ov#1{\overline{#1}}

\def\vb#1{\mbox{\boldmath$#1$}}

\def\wh#1{\widehat{#1}}
\def\bdot{\,\vb{\cdot}\,}
\def\btimes{\,\vb{\times}\,}

\def\bhat{\wh{{\sf b}}}
\def\cal#1{\mathcal{#1}}

\newcommand{\bc}{\begin{center}}
\newcommand{\ec}{\end{center}}
\newcommand{\bt}{\begin{tabbing}}
\newcommand{\et}{\end{tabbing}} 
\newcommand{\be}{\begin{eqnarray*}}
\newcommand{\ee}{\end{eqnarray*}}
\newcommand{\bs}{\begin{slide}}
\newcommand{\es}{\end{slide}}

\begin{document}

\title{Equivalent elastica knots}

\author{Alain J.~Brizard}
\affiliation{Department of Physics, Saint Michael's College, Colchester, VT 05439, USA} 

\author{David Pfefferl\'{e}}
\affiliation{The University of Western Australia, 35 Stirling Highway, Crawley WA 6009, AUSTRALIA}

\begin{abstract}
The problem of an elastica knot in three-dimensional space is solved explicitly by expressing the Frenet-Serret curvature and torsion of the knot in terms of the Weierstrass and Jacobi elliptic functions. This solution is obtained by variational methods and is derived by minimizing of the squared-curvature energy integral. In the present work, an equivalency is established between pairs of Jacobi elliptic solutions that are described by the same values for curvature and torsion functionals.
\end{abstract}

%\pacs{52.30.Gz, 52.65.Tt}

\maketitle

\section{Introduction}

Elastica knots \cite{Langer_Singer_1984,LS_1984,Singer_2008} are three-dimensional closed curves that minimize the constrained curvature functional
\begin{equation}
{\cal F}_{\Lambda}[{\bf r}] = \frac{1}{2}\;\int_{a}^{b} \left[ |{\bf r}^{\prime\prime}(s)|^{2} \;+\frac{}{} \Lambda(s)\;\left(|{\bf r}^{\prime}(s)|^{2} - 1 \right) \right] ds,
\label{eq:F_Lambda}
\end{equation}
where the curve ${\bf r}(s)$ is parameterized by the length element $s$ along the curve, and the function $\Lambda(s)$ serves as a Lagrange multiplier associated with the constraint 
$|{\bf r}^{\prime}(s)|^{2} = 1$ (since $ds^{2} \equiv |d{\bf r}|^{2}$).

Here, the three-dimensional curve ${\bf r}(s)$ satisfies the Frenet-Serret equations
\begin{equation}
\left( \begin{array}{c}
d\,\wh{\sf t}/ds \\
d\,\wh{\sf n}/ds \\
d\,\wh{\sf b}/ds
\end{array} \right) \;=\; \left( \begin{array}{ccc}
0 & \kappa & 0 \\
-\,\kappa & 0 & \tau \\
0 & -\,\tau & 0
\end{array} \right) \cdot \left( \begin{array}{c}
\wh{\sf t} \\
\wh{\sf n} \\
\wh{\sf b}
\end{array} \right),
\label{eq:FS}
\end{equation}
where the tangent unit vector $\wh{\sf t} \equiv {\bf r}^{\prime}$, the normal unit vector $\wh{\sf n}$, and the binormal unit vector $\wh{\sf b} \equiv \wh{\sf t}\btimes\wh{\sf n}$ form the Frenet-Serret unit-vector triad $(\wh{\sf t},\wh{\sf n},\wh{\sf b})$, while $\kappa$ and $\tau$ denote the curvature and the torsion of the curve, respectively. Using the Frenet-Serret formulas \eqref{eq:FS}, we also derive the following expressions
\begin{equation}
\left. \begin{array}{rcl}
{\bf r}^{\prime} & = & \wh{\sf t} \\
{\bf r}^{\prime\prime} & = & \kappa\;\wh{\sf n} \\
{\bf r}^{\prime\prime\prime} & = & \kappa^{\prime}\;\wh{\sf n} \;+\; \kappa\,\left( \tau\;\wh{\sf b} \;-\; \kappa\;\wh{\sf t}\right)
\end{array} \right\},
\label{eq:r_primes}
\end{equation}
from which we obtain the definitions for the Frenet-Serret curvature $\kappa(s) \equiv |{\bf r}^{\prime\prime}|$ and the Frenet-Serret torsion $\tau(s) \equiv \kappa^{-2}\,({\bf r}^{\prime}\btimes
{\bf r}^{\prime\prime}\bdot{\bf r}^{\prime\prime\prime})$. We note that the torsion may be positive, negative, or zero (i.e., when the curve lies on a two-dimensional plane).

The Euler equation for the curve ${\bf r}(s)$ is obtained from the first variation of the curvature functional \eqref{eq:F_Lambda}:
\begin{eqnarray}
\delta{\cal F}_{\Lambda} & \equiv & \left(\frac{d}{d\epsilon}{\cal F}_{\Lambda}[{\bf r} + \epsilon\,\delta{\bf r}]\right)_{\epsilon = 0}  \label{eq:delta_F} \\
 & = & \int_{a}^{b}\left[ {\bf r}^{\prime\prime}\bdot\delta{\bf r}^{\prime\prime} +\frac{}{} \Lambda\;{\bf r}^{\prime}\bdot\delta{\bf r}^{\prime}\right] ds = \int_{a}^{b} \delta{\bf r}\bdot\frac{d{\bf W}}{ds}ds,
\nonumber
\end{eqnarray}
where the variation $\delta{\bf r}$ and its first derivative $\delta{\bf r}^{\prime}$ are assumed to vanish at the end points $s = a$ and $s = b$, and the vector
\begin{eqnarray}
{\bf W}(s) & \equiv & {\bf r}^{\prime\prime\prime} - \Lambda\;{\bf r}^{\prime} = \kappa^{\prime}\;\wh{\sf n} + \kappa\,\tau\;\wh{\sf b} - \left(\kappa^{2} + \Lambda\right)\;\wh{\sf t}
 \label{eq:W_def}
 \end{eqnarray}
is written in terms of Eq.~\eqref{eq:r_primes} and the Lagrange multiplier $\Lambda(s)$. When the first variation \eqref{eq:delta_F} vanishes for arbitrary variations $\delta{\bf r}$ (subject to vanishing boundary conditions), we obtain the Euler equation relating the curvature $\kappa$ and the torsion $\tau$ for the curve ${\bf r}(s)$:
 \begin{eqnarray}
 0 = \frac{d{\bf W}}{ds} & = & -\;\left( 3\,\kappa\,\kappa^{\prime} \;+\frac{}{} \Lambda^{\prime}\right)\wh{\sf t} + \left( 2\,\kappa^{\prime}\;\tau \;+\frac{}{} \kappa\;\tau^{\prime}\right) \wh{\sf b} \nonumber \\
  &  &+\; \left[\kappa^{\prime\prime} - \kappa \left( \kappa^{2} \;+\frac{}{}\tau^{2} + \Lambda\right) \right] \wh{\sf n}.
 \label{eq:Euler-W}
 \end{eqnarray}
 The $\wh{\sf t}$-component of Eq.~\eqref{eq:Euler-W} yields the conservation law $(\frac{3}{2}\,\kappa^{2} + \Lambda)^{\prime} = 0$, from which we obtain a solution for the Lagrange multiplier
 \begin{equation}
 \Lambda(s) \;\equiv\; -\;\frac{3}{2}\;\kappa^{2}(s) \;+\; \frac{1}{2}\,\lambda\,k_{0}^{2},
 \label{eq:Lambda_eq}
 \end{equation}
 where $\lambda$ denotes a dimensionless constant of integration (initially assumed to be $-\infty < \lambda < \infty$) and the curvature parameter $k_{0}$ is defined as $k_{0} \equiv \kappa(0)$. The $\wh{\sf b}$-component of 
 Eq.~\eqref{eq:Euler-W} yields the conservation law $(\kappa^{2}\,\tau)^{\prime} = 0$, from which we obtain the torsion constraint
 \begin{equation}
 \kappa^{2}(s)\;\tau(s) \;=\; {\bf r}^{\prime}\btimes{\bf r}^{\prime\prime}\bdot{\bf r}^{\prime\prime\prime} \;\equiv\; k_{0}^{2}\,\tau_{0},
 \label{eq:tau_eq}
 \end{equation}
 where the torsion parameter $\tau_{0} $ is defined as $\tau_{0} \equiv \tau(0)$. Substituting Eqs.~\eqref{eq:Lambda_eq}-\eqref{eq:tau_eq} into Eq.~\eqref{eq:W_def}, it becomes a function of $\kappa$ and  $\kappa^{\prime}$:
 \begin{equation}
{\bf W} \;=\; \kappa^{\prime}\;\wh{\sf n} \;+\; k_{0}^{2}\tau_{0}\,\kappa^{-1}\;\wh{\sf b} \;+\; \frac{1}{2}\,\left(\kappa^{2} \;-\frac{}{}\lambda\,k_{0}^{2} \right)\wh{\sf t}.
 \label{eq:W_const}
 \end{equation}
 Lastly, the $\wh{\sf n}$-component of Eq.~\eqref{eq:Euler-W} yields the curvature second-order ordinary differential equation
 \begin{eqnarray}
 \kappa^{\prime\prime} & = & -\;\frac{1}{2}\,\kappa^{3} \;+\; k_{0}^{4}\tau_{0}^{2}\;\kappa^{-3} \;+\; \frac{1}{2}\;\lambda\,k_{0}^{2}\kappa,
  \label{eq:kappa_pp}
  \end{eqnarray}
  where we inserted the relations \eqref{eq:Lambda_eq}-\eqref{eq:tau_eq}.
   
 \section{\label{sec:curvature}Curvature Equation}
 
 In this Section, we solve the curvature second-order ordinary differential equation \eqref{eq:kappa_pp} in terms of the Weierstrass elliptic function \cite{NIST_Weierstrass,Lawden,Brizard_2015} and the Jacobi elliptic function \cite{NIST_Jacobi}.  We begin with the derivation of the general solution of Eq.~\eqref{eq:kappa_pp} expressed in terms of the Weierstrass elliptic function. Next, using standard relations among elliptic functions, we derive the Jacobi elliptic solution from the Weierstrass solution.
 
 Before proceeding with our solution, however, we transform Eq.~\eqref{eq:kappa_pp} as follows. First, we multiply Eq.~\eqref{eq:kappa_pp} by $\kappa^{\prime}$ and integrate it to obtain 
 \begin{eqnarray}
(\kappa^{\prime})^{2} & = & -\,\frac{1}{4} \left( \kappa^{4} \;-\; k_{0}^{4}\right) \;-\; k_{0}^{2}\tau_{0}^{2}\;\left(\frac{k_{0}^{2}}{\kappa^{2}} \;-\; 1\right) \nonumber \\
 &  &+\; \frac{1}{2}\;\lambda\,k_{0}^{2}\;\left( \kappa^{2} \;-\; k_{0}^{2} \right),
 \label{eq:K_prime}
 \end{eqnarray}
 where we used the initial conditions $\kappa(0) \equiv k_{0}$ and $\kappa^{\prime}(0) \equiv 0$. Next, we multiply Eq.~\eqref{eq:K_prime} by $4\kappa^{2}$ to obtain the 
 squared-curvature equation
 \begin{eqnarray}
 \left[\left(\kappa^{2}\right)^{\prime}\right]^{2} & = & -\,\kappa^{6} + 2\lambda\,k_{0}^{2}\kappa^{4}  \;-\; \nu^{2}\,k_{0}^{6} \nonumber \\
  &  &+\; k_{0}^{4}\,\kappa^{2} \left[ (1 - 2\,\lambda) +\frac{}{} \nu^{2} \right],
 \label{eq:K2_prime}
 \end{eqnarray}
where we introduced the dimensionless real-valued torsion parameter $\nu \equiv 2\,\tau_{0}/k_{0}$, which can be positive, negative, or zero.
 
 \subsection{Weierstrass elliptic solution}
 
Our goal is now to transform Eq.~\eqref{eq:K2_prime} into the standard Weierstrass form \cite{NIST_Weierstrass,Lawden}
 \begin{eqnarray}
 \left(\frac{d\wp(z)}{dz}\right)^{2} & = & 4\;\wp^{3}(z) \;-\; g_{2}\;\wp(z) \;-\; g_{3}  \label{eq:WP_form} \\
  & \equiv & 4\left[\wp(z) - {\sf e}_{1}\right]\left[\wp(z) - {\sf e}_{2}\right]\left[\wp(z) - {\sf e}_{3}\right],
\nonumber
 \end{eqnarray}
 where $\wp(z;g_{2},g_{3})$ denotes the Weierstrass elliptic function (which is an even-parity doubly-periodic function of its argument $z$) and the cubic roots ${\sf e}_{k} = ({\sf e}_{1},{\sf e}_{2},{\sf e}_{3})$ satisfy the relation ${\sf e}_{1} + {\sf e}_{2} + {\sf e}_{3} = 0$ as well as the ordering
 \begin{equation}
 {\sf e}_{3} \;\leq\; {\sf e}_{2} \;\leq\;  {\sf e}_{1}
 \label{eq:e_123}
 \end{equation} 
when the roots are real. The invariants $(g_{2}, g_{3}, \Delta)$ of $\wp(z;g_{2},g_{3})$ are
 \begin{equation}
 \left. \begin{array}{rcl}
 g_{2} & \equiv & 2\left({\sf e}_{1}^{2} + {\sf e}_{2}^{2} + {\sf e}_{3}^{2}\right) \\
 g_{3} & \equiv & 4\,{\sf e}_{1}{\sf e}_{2}{\sf e}_{3} \\
 \Delta & \equiv & g_{2}^{3} \;-\; 27 \;g_{3}^{2} \\
  & = & 16\,({\sf e}_{1} - {\sf e}_{2})^{2}\,({\sf e}_{2} - {\sf e}_{3})^{2}\,({\sf e}_{1} - {\sf e}_{3})^{2} 
 \end{array} \right\}.
 \label{eq:g23_Delta}
 \end{equation}
The half-periods $\omega_{k}(g_{2},g_{3}) = (\omega_{1},\omega_{2},\omega_{3})$, which also depend on the sign of the modular discriminant $\Delta$ \cite{Brizard_2015}, satisfy the periodicity conditions 
\[ \wp(z + 2\,\omega_{k}; g_{2},g_{3}) \;=\; \wp(z;g_{2},g_{3}), \]
with the definitions $\wp(\omega_{k};g_{2},g_{3}) = {\sf e}_{k}$ and $\wp^{\prime}(\omega_{k};g_{2},g_{3}) = 0$. When the three roots $ ({\sf e}_{1},{\sf e}_{2},{\sf e}_{3})$ are real, $\omega_{1}$ is real, $\omega_{3}$ is imaginary, and $\omega_{2} \equiv -\,\omega_{1} - \omega_{3}$ is complex.
 
 For the purpose of transforming Eq.~\eqref{eq:K2_prime} into the standard Weierstrass form \eqref{eq:WP_form}, we introduce the transformation
 \begin{equation}
 \kappa^{2}(s) \;\equiv\; \frac{k_{0}^{2}}{q_{0}}\;q(\varphi),
 \label{eq:kappa_q}
 \end{equation}
 where $\varphi(s) \equiv i\;k_{0}s/(2\sqrt{q_{0}}) + \varphi_{0}$, with $\varphi_{0}$ chosen so that $q_{0} \equiv q(\varphi_{0})$ denotes a real-valued scale parameter. Substituting the transformation \eqref{eq:kappa_q} into Eq.~\eqref{eq:K2_prime} yields
 \begin{eqnarray}
 [q^{\prime}(\varphi)]^{2} & = & 4\,q_{0}^{3}\;\nu^{2} \;-\; 4\,q_{0}^{2}q \left( 1 - 2\lambda \;+\frac{}{} \nu^{2} \right)  \nonumber \\
  &  &-\; 8\,\lambda\,q_{0}q^{2} \;+\; 4\,q^{3}  \;\equiv\; Q(q).
 \label{eq:q_prime}
 \end{eqnarray}
We note that the right side of Eq.~\eqref{eq:q_prime} is a cubic polynomial $Q(q)$, which is not yet in the Weierstrass form \eqref{eq:WP_form}.

\begin{figure}
\epsfysize=2in
\epsfbox{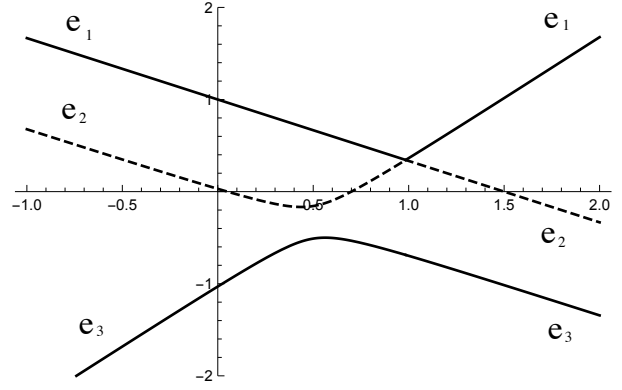}
\caption{Plots of the normalized cubic roots $\ov{\sf e}_{k}(\lambda,\nu) \equiv {\sf e}_{k}/q_{0}$ as functions of $\lambda$ for $\nu = \nu_{0}/2 = (4\sqrt{2})^{-1}$: $\ov{\sf e}_{1}$ (top solid curve) $\geq\;\ov{\sf e}_{2}$ (dashed curve) $> \ov{\sf e}_{3}$ (bottom solid curve). The two roots $\ov{\sf e}_{1} = \ov{\sf e}_{2}$ merge at $\lambda = \lambda_{\Delta} = 1 - \nu^{2}/2$, where $\Delta(\lambda,\nu)$ vanishes.}
\label{fig:ek_W}
\end{figure}

In order to bring Eq.~\eqref{eq:q_prime} into the Weierstrass form \eqref{eq:WP_form}, we consider the uniform translation $q(\varphi) = \chi(q_{0}) + \wp(\varphi)$, where $\wp(\varphi_{0}) \equiv q_{0} - 
\chi(q_{0})$ and $\wp^{\prime}(\varphi_{0}) = 0$, so that Eq.~\eqref{eq:q_prime} becomes
 \begin{equation}
 (\wp^{\prime})^2 \;=\; 4\,\wp^{3} + \frac{1}{2}\,Q^{\prime\prime}(\chi)\;\wp^{2} + Q^{\prime}(\chi)\;\wp + Q(\chi). 
 \end{equation}
 Since the $\wp^{2}$-term is absent in Eq.~\eqref{eq:WP_form}, we must choose $Q^{\prime\prime}(\chi) = 24\,\chi - 16\,\lambda\,q_{0} \equiv 0$, which yields the constant
 \begin{equation}
 \chi(q_{0}) \;=\; \frac{2}{3}\,\lambda\,q_{0}.
 \label{eq:lambda_q0}
 \end{equation}
Hence, the invariant functions $g_{2} \equiv -\,Q^{\prime}(\chi)$ and $g_{3} \equiv -\,Q(\chi)$ are now expressed as
 \begin{eqnarray}
 g_{2}(\lambda,\nu,q_{0}) & = &  \frac{4}{3}\,\left( 3 \;-\frac{}{} 6\,\lambda + 4\,\lambda^{2} + 3\,\nu^{2}\right) q_{0}^{2} \nonumber \\
  & \equiv & \ov{g}_{2}(\lambda,\nu)\,q_{0}^{2}, \label{eq:g2} \\
 g_{3}(\lambda,\nu,q_{0}) & = &  \frac{8}{27}\,\left(\lambda \;-\; \frac{3}{2}\right)\;\left( 8 \lambda^{2} \;-\frac{}{} 6\,\lambda + 9\,\nu^{2}\right) q_{0}^{3} \nonumber \\
  & \equiv & \ov{g}_{3}(\lambda,\nu)\,q_{0}^{3}. \label{eq:g3}
 \end{eqnarray} 
The modular discriminant $\Delta \equiv g_{2}^{3} - 27\,g_{3}^{2}$, on the other hand, is
 \begin{eqnarray}
 \Delta(\lambda,\nu,q_{0}) & = &  64\left( \lambda \;-\frac{}{} \lambda_{\Delta}(\nu) \right)^{2} \delta^{2}(\lambda,\nu)\;q_{0}^{6} \nonumber \\
  & \equiv & \ov{\Delta}(\lambda,\nu)\, q_{0}^{6},
 \label{eq:Delta_general}
 \end{eqnarray}
where 
\begin{equation}
\left. \begin{array}{rcl}
\lambda_{\Delta}(\nu) & = & 1 - \nu^{2}/2 \\
 &  & \\
\delta^{2}(\lambda,\nu) & = & (1 - 2\lambda)^{2} + 4\,\nu^{2}
\end{array} \right\}. 
\end{equation}
We note that the invariant function $g_{2}$ does not vanish and, hence, the cubic roots are real and satisfy the ordering \eqref{eq:e_123}. The invariant function $g_{3}$ vanishes when one of the roots vanishes, e.g., when $\lambda = 3/2$ in Eq.~\eqref{eq:g3}; we note that $g_{3}$ does not vanish anywhere else if $\nu > \nu_{0} \equiv 1/\sqrt{8}$. The modular discriminant $\Delta$, on the other hand, is non-negative (since the roots are real) and vanishes when two roots merge, i.e., when $\lambda = \lambda_{\Delta}$ (see Fig.~\ref{fig:ek_W}). When we evaluate Eq.~\eqref{eq:kappa_pp} at $s = 0$, we find
\begin{equation}
\kappa_{0}^{\prime\prime} \;=\; \frac{k_{0}^{3}}{2} \left(\lambda \;-\frac{}{} \lambda_{\Delta}\right),
\label{eq:kappa_dp}
\end{equation}
which means that $k_{0}$ is a maximum $(\kappa_{0}^{\prime\prime} < 0)$ when $\lambda < \lambda_{\Delta}$ and $k_{0}$ is a minimum $(\kappa_{0}^{\prime\prime} > 0)$ when $\lambda > \lambda_{\Delta}$.
We also note that the invariant functions \eqref{eq:g2}-\eqref{eq:Delta_general} are homogeneous functions of the scale parameter $q_{0}$; this important remark will form the basis of a new parametrization introduced in Sec.~\ref{sec:W}.
  
The squared-curvature equation \eqref{eq:kappa_q} is, therefore, solved in terms of the Weierstrass elliptic function as
 \begin{eqnarray}
 \kappa^{2}(s) & = & k_{0}^{2} \left[ \frac{2}{3}\,\lambda + q_{0}^{-1}\;\wp\left(\varphi; g_{2}, g_{3}\right) \right]  \label{eq:kappa_PWJ} \\
  & = & \frac{k_{0}^{2}}{q_{0}} \left[ q_{0} \;+\frac{}{} \wp\left(i\xi + \omega_{a}; g_{2}, g_{3}\right) - {\sf e}_{a} \right],
\nonumber
 \end{eqnarray}
 where $\varphi(s) \equiv i\xi(s) + \omega_{a}$, with 
 \begin{equation}
 \xi(s) \;=\; k_{0}s/(2\sqrt{q_{0}}) \;\equiv\; \ov{\xi}(s)/\sqrt{q_{0}}, 
 \label{eq:xi_def}
 \end{equation}
 and
 \begin{equation}
 {\sf e}_{a} \;\equiv\; \left(1 - \frac{2}{3}\,\lambda\right)\,q_{0} \;\equiv\; \ov{\sf e}_{a}(\lambda,\nu)\,q_{0}
 \label{eq:e1_chi}
 \end{equation}
denotes one of the three cubic roots ${\sf e}_{k} = ({\sf e}_{1},{\sf e}_{2},{\sf e}_{3})$, with the half-period $\omega_{a}$ defined from the identity $\wp(\omega_{a};g_{2},g_{3}) \equiv 
{\sf e}_{a}$.  As indicated above, the cubic roots $({\sf e}_{1},{\sf e}_{2},{\sf e}_{3})$ of the Weierstrass elliptic function form an ordered set \eqref{eq:e_123} when the roots are real. From Eqs.~\eqref{eq:g23_Delta} and \eqref{eq:g2}-\eqref{eq:g3}, and using Eq.~\eqref{eq:e1_chi}, the other two cubic roots are
 \begin{eqnarray}
 {\sf e}_{b,c} & = & -\,\frac{1}{2}\,{\sf e}_{a} \;\pm\; \frac{q_{0}}{2}\;\delta(\lambda,\nu) \;\equiv\; \ov{\sf e}_{b,c}(\lambda,\nu)\,q_{0},
 \label{eq:e23_chi}
 \end{eqnarray}
which are also real. 
 
Figure \ref{fig:ek_W} shows the normalized cubic roots $\ov{\sf e}_{k}(\lambda,\nu) \equiv {\sf e}_{k}/q_{0}$, defined by Eqs.~\eqref{eq:e1_chi}-\eqref{eq:e23_chi}, as functions of $\lambda$ for $\nu = \nu_{0}/2 = (4\sqrt{2})^{-1}$ (i.e., $g_{3}$ vanishes three times when ${\sf e}_{2}$ vanishes). The cubic root $\ov{\sf e}_{1}$ (shown as the top solid curve) is chosen to be larger than the cubic root $\ov{\sf e}_{2}$ (shown as a dashed curve), and the two roots $\ov{\sf e}_{1} =  \ov{\sf e}_{2}$ merge at $\lambda_{\Delta} \equiv 1 - \nu^{2}/2$ (i.e., when $\Delta = 0$). The third cubic root $\ov{\sf e}_{3}$ (shown as the bottom solid curve) always satisfies the ordering \eqref{eq:e_123} when $\nu \neq 0$. 

Lastly, if we now require that the solution \eqref{eq:kappa_PWJ} be periodic in $0 \leq s \leq S(\lambda,\nu,k_{0})$, the condition $\kappa^{2}(S) = k_{0}^{2}$ yields the definition
\begin{equation}
k_{0}\,S \equiv 4\,\sqrt{q_{0}}\;|\omega_{3}(\lambda,\nu,q_{0})| \;=\; 4\,|\ov{\omega}_{3}(\lambda,\nu)|,
\label{eq:S_def}
\end{equation}
where the half-period $\omega_{3} \equiv i\;|\omega_{3}|$ is purely imaginary since the roots are real. Here, we note that the half-periods $\ov{\omega}_{a} \equiv \omega_{a}\,\sqrt{q_{0}}$ and the period \eqref{eq:S_def} are independent of the scale parameter $q_{0}$.
 
 \subsection{Weierstrass cubic roots and half-periods}

 From Fig.~\ref{fig:ek_W}, we identify the cubic root \eqref{eq:e1_chi} as
 \begin{equation}
 \left(1 - \frac{2}{3}\,\lambda\right)\,q_{0} = {\sf e}_{a} \equiv \left\{ \begin{array}{lcr}
 {\sf e}_{1}^{-} &  & (\lambda < \lambda_{\Delta}) \\
  &  & \\
 {\sf e}_{2}^{+} &  & (\lambda > \lambda_{\Delta})
 \end{array} \right.
 \label{eq:cubic_a}
 \end{equation}
 where the $\pm$ notation is based on the sign of $\lambda - \lambda_{\Delta}$ (i.e., the sign of $\kappa_{0}^{\prime\prime}$). The remaining cubic roots \eqref{eq:e23_chi} are identified as
 \begin{equation}
-\,\frac{1}{2}\,{\sf e}_{a}  + \frac{q_{0}}{2}\;\delta = {\sf e}_{b} \equiv  \left\{ \begin{array}{lcr}
 {\sf e}_{2}^{-} &  & (\lambda < \lambda_{\Delta}) \\
  &  & \\
 {\sf e}_{1}^{+} &  & (\lambda > \lambda_{\Delta})
 \end{array} \right.
  \label{eq:cubic_b}
 \end{equation}
 and
 \begin{equation}
-\,\frac{1}{2}\,{\sf e}_{a} \;-\; \frac{q_{0}}{2}\;\delta(\lambda,\nu) \;\equiv\;  {\sf e}_{3}^{\pm},
  \label{eq:cubic_c}
 \end{equation}
 for all values of $(\lambda,\nu)$. Using Eqs.~\eqref{eq:cubic_a}-\eqref{eq:cubic_b}, we also obtain the following expression for $\nu^{2}$:
 \begin{eqnarray}
 \nu^{2} & = &\left({\sf e}_{b} - {\sf e}_{a} \frac{}{}+ q_{0}\right) \left({\sf e}_{a} -{\sf e}_{3} \frac{}{}- q_{0}\right)/q_{0}^{2} \nonumber \\
 & \equiv & \left(\ov{\sf e}_{b} - \ov{\sf e}_{a} \frac{}{}+ 1\right) \left(\ov{\sf e}_{a} - \ov{\sf e}_{3} \frac{}{}- 1\right),
 \label{eq:nu2_def}
 \end{eqnarray}
 where we made use of the homogeneity of the roots as functions of $q_{0}$. The torsionless case $\nu = 0$ yields the roots
 \begin{equation}
 \left. \begin{array}{rcl}
 {\sf e}_{a} & = &  (1 - 2\,\lambda/3)\,q_{0} \\
 {\sf e}_{b} & = & (-2\,\lambda/3)\,q_{0} \\
 {\sf e}_{c} & = & (-1 + 4\,\lambda/3)\,q_{0}
\end{array} \right\},
\label{eq:torsion_0}
\end{equation} 
where ${\sf e}_{c} \leq {\sf e}_{b} < {\sf e}_{a}$ for $\lambda \leq \frac{1}{2}$, ${\sf e}_{b} < {\sf e}_{c} \leq {\sf e}_{a}$ for $\frac{1}{2} < \lambda \leq 1$, and ${\sf e}_{b} < {\sf e}_{a} < {\sf e}_{c}$ for $\lambda > 1$.
 
\begin{figure}
\epsfysize=2in
\epsfbox{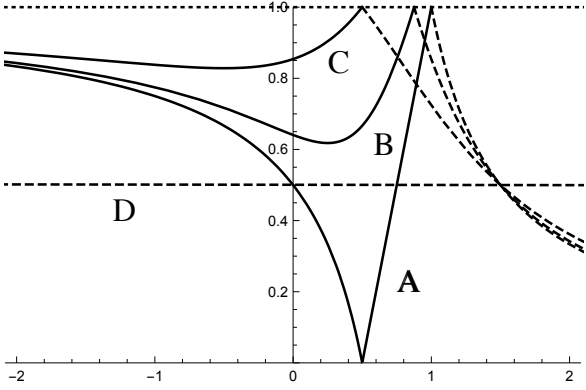}
\caption{Plot of the modulus $0 \leq p(\lambda,\nu) \leq 1$ as a function of $\lambda$ in the ranges $\lambda < \lambda_{\Delta}$ (solid) and $\lambda > \lambda_{\Delta}$ (dashed) for various values of $\nu$: 
$\nu = 0$ (curve A), $\nu = \frac{1}{2}$ (curve B), $\nu = 1$ (curve C), and $\nu \gg 1$ (curve D). The modulus $p(\lambda,\nu) = 1$ when $\lambda = \lambda_{\Delta}$ and $p(\lambda,\nu) = \frac{1}{2}$ for all values of $\nu$ at $\lambda = \frac{3}{2}$.}
\label{fig:m_W}
\end{figure}

 The root identifications \eqref{eq:cubic_a}-\eqref{eq:cubic_c} allow us to define the Jacobi elliptic modulus $p$ and the complementary modulus $p^{\prime} \equiv 1 - p$ for all values of $(\lambda,\nu)$:
 \begin{equation}
 p(\lambda,\nu) \equiv \frac{{\sf e}_{2} - {\sf e}_{3}}{{\sf e}_{1} - {\sf e}_{3}} \;\; {\rm and}\;\; p^{\prime}(\lambda,\nu) \equiv \frac{{\sf e}_{1} - {\sf e}_{2}}{{\sf e}_{1} - {\sf e}_{3}},
 \label{eq:p_modulus}
 \end{equation}
 where the independence on the scale parameter $q_{0}$ follows from the homogeneity of the cubic roots ${\sf e}_{k} \equiv q_{0}\,\ov{\sf e}_{k}(\lambda,\nu)$. From Fig.~\ref{fig:m_W}, we note that the boundary value $\lambda = \lambda_{\Delta}$, we find $p^{\pm}(\lambda_{\Delta},\nu) = 1$, while we find $p^{+}(\frac{3}{2},\nu) = \frac{1}{2}$ at $\lambda = \frac{3}{2}$. We note that, by definition, both $p$ and $p^{\prime} = 1 - p$ lie in the closed interval $[0,1]$.
 
\begin{figure}
\epsfysize=2in
\epsfbox{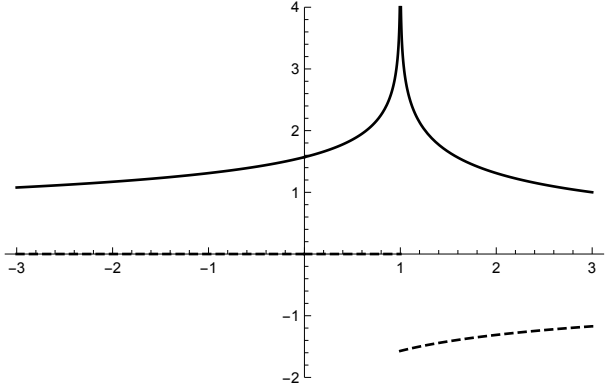}
\caption{Plots of the real (solid) and imaginary (dashed) parts of the elliptic integral ${\sf K}(p)$ in the range $-3 \leq p \leq 3$.}
\label{fig:Elliptic_K}
\end{figure}

In Eq.~\eqref{eq:W_J} below, the two-parameter real half-period of the Weierstrass elliptic function is
\begin{eqnarray}
\ov{\omega}_{1}(\lambda,\nu) & = & \int_{{\sf e}_{1}}^{\infty}\frac{\sqrt{q_{0}}\;dz}{\sqrt{4\,(z - {\sf e}_{1})\,(z - {\sf e}_{2})\,(z - {\sf e}_{3})}} \nonumber  \\
 & = & \int_{\sqrt{{\sf e}_{1} - {\sf e}_{3}}}^{\infty}\frac{\sqrt{q_{0}}\;ds}{\sqrt{[s^{2} - ({\sf e}_{1} - {\sf e}_{3})]\,[s^{2} - ({\sf e}_{2} - {\sf e}_{3})]}} \nonumber \\
 & = & \int_{0}^{\pi/2}\frac{\sqrt{q_{0}}\;d\theta}{\sqrt{ ({\sf e}_{1} - {\sf e}_{3}) \;-\; ({\sf e}_{2} - {\sf e}_{3})\;\sin^{2}\theta}} \nonumber \\
  & \equiv & \frac{{\sf K}(p)}{\sqrt{\ov{\sf e}_{1} - \ov{\sf e}_{3}}},
\label{eq:omega1_p}
\end{eqnarray}
and the two-parameter imaginary half-period is
\begin{eqnarray}
\ov{\omega}_{3}(\lambda,\nu) & = &\int_{-\infty}^{{\sf e}_{3}}\frac{ \pm\;i\;\sqrt{q_{0}}\;dz}{\sqrt{4\,({\sf e}_{1} - z)\,({\sf e}_{2} - z)\,({\sf e}_{3} - z)}} \nonumber \\
 & = & \int_{\sqrt{{\sf e}_{1} - {\sf e}_{3}}}^{\infty}\frac{\pm\; i\;\sqrt{q_{0}}\;ds}{\sqrt{[s^{2} - ({\sf e}_{1} - {\sf e}_{3})]\,[s^{2} - ({\sf e}_{1} - {\sf e}_{2})]}} \nonumber \\
 & = & \int_{0}^{\pi/2}\frac{\pm\;i\;\sqrt{q_{0}}\;d\theta}{\sqrt{ ({\sf e}_{1} - {\sf e}_{3}) \;-\; ({\sf e}_{1} - {\sf e}_{2})\;\sin^{2}\theta}} \nonumber \\
 & \equiv & \pm\;i\;\frac{{\sf K}(1-p)}{\sqrt{\ov{\sf e}_{1} - \ov{\sf e}_{3}}},
\label{eq:omega3_p}
\end{eqnarray} 
where the sign $\pm$ is determined from the sign of $g_{3} = 4\,{\sf e}_{1}{\sf e}_{2}{\sf e}_{3}$. Here, the complete elliptic integral of the first kind 
\begin{equation}
{\sf K}(p) \;\equiv\; \int_{0}^{\pi/2}\frac{d\theta}{\sqrt{1 - p\,\sin^{2}\theta}}
\label{eq:K_def}
\end{equation}
is defined in terms of the {\sf Mathematica} notation, which differs from the standard definition ${\sf K}(m)$ defined with $p = m^{2}$ in  the integrand of 
Eq.~\eqref{eq:K_def}. The integral \eqref{eq:K_def} is defined for the Jacobi modulus in the range $0 \leq p < 1$, and its special values are ${\sf K}(0) = \pi/2$ and ${\sf K}(p) \rightarrow \infty$ as $p \rightarrow 1$. In the range $p \leq 0$, we easily find
\begin{equation}
{\sf K}(p) \;=\; {\sf K}(-p/p^{\prime})/\sqrt{p^{\prime}},
\label{eq:K_neg}
\end{equation}
where $p^{\prime} \equiv 1 - p > -\,p \geq 0$, while in the range $p \geq 1$, we find
\begin{equation}
{\sf K}(p) \;=\; \left[{\sf K}(1/p) \;-\frac{}{} i\;{\sf K}(1 - 1/p)\right]/\sqrt{p},
\label{eq:K_plus}
\end{equation}
The complete elliptic integral of the first kind \eqref{eq:K_def} is shown in Fig.~\ref{fig:Elliptic_K} in the range $-3 \leq p \leq 3$, where the extended formulas \eqref{eq:K_neg} and \eqref{eq:K_plus} are used.
 
Lastly, the two half-periods \eqref{eq:omega1_p}-\eqref{eq:omega3_p} can  be used to construct the third (complex-valued) half-period \cite{NIST_Weierstrass}
\begin{equation}
\omega_{2} \;\equiv\; -\;\omega_{1} \;-\; \omega_{3}. 
\label{eq:omega2_p}
\end{equation}
The three half-periods $(\omega_{1},\omega_{2},\omega_{3})$ will appear extensively in what follows.

\subsection{Jacobi elliptic solution}

We can transform the Weierstrass elliptic solution \eqref{eq:kappa_PWJ} into a Jacobi elliptic solution by using the appropriate relations to the Jacobi elliptic functions
\cite{NIST_Jacobi}:
 \begin{equation}
 \left. \begin{array}{l}
 \wp(iz+\omega_{1}) - {\sf e}_{1} = -({\sf e}_{1} - {\sf e}_{2})\;{\rm sn}^{2}(u|p^{\prime}) \\
 \\
 \wp(iz+\omega_{2}) - {\sf e}_{2} = ({\sf e}_{1} - {\sf e}_{2})\;p\,{\rm sd}^{2}(u|p^{\prime})
\end{array} \right\},
 \label{eq:W_J}
 \end{equation}
where $u \equiv z\sqrt{{\sf e}_{1} - {\sf e}_{3}}$ and ${\rm sn}^{2}(u|p^{\prime})$ and ${\rm sd}^{2}(u|p^{\prime}) \equiv {\rm sn}^{2}(u|p^{\prime})/{\rm dn}^{2}(u|p^{\prime})$ are Jacobi elliptic functions with real and imaginary half-periods ${\sf K}(p^{\prime})$ and $\pm\,i\,{\sf K}(p)$, respectively. Here, we note that we use the {\sf Mathematica} convention ${\rm sn}(\xi,k) = {\rm sn}(\xi|k^{2})$, with standard formulas \cite{NIST_Jacobi} expressed in terms of ${\rm sn}(\xi,k)$. The {\sf Mathematica} convention, therefore, immediately satisfies the symmetry ${\rm sn}(\xi,-\,k) = {\rm sn}(\xi,k)$, while we also find the transformation
\begin{eqnarray}
{\rm sn}(\xi, i\,k) & = & {\rm sn}\left(\xi\left|-\,k^{2}\right.\right) 
\label{eq:sn_transform} \\
 & = & \frac{1}{\sqrt{1+k^{2}}}\,{\rm sd}\left(\xi\sqrt{1+k^{2}},\;k/\sqrt{1+k^{2}}\right)
 \nonumber \\
 & = & \frac{1}{\sqrt{1+k^{2}}}\,{\rm sd}\left(\xi\sqrt{1+k^{2}}\left|\frac{}{}k^{2}/(1+k^{2})\right.\right),
 \nonumber 
\end{eqnarray}
where $k$ is any real number.

The general Jacobi solutions of the squared-curvature equation \eqref{eq:K2_prime} are
\begin{equation}
\kappa_{-}^{2}(s) = \frac{k_{0}^{2}}{q_{0}} \left[ q_{0} \;-\frac{}{} \left({\sf e}_{1}^{-} - {\sf e}_{2}^{-}\right) {\rm sn}^{2}\left(u^{-}\,|\,p^{\prime -}\right) \right],
\label{eq:kappa2_JI}
\end{equation}
where $\lambda < \lambda_{\Delta}$ and $u^{-} \equiv \xi\sqrt{{\sf e}_{1}^{-} - {\sf e}_{3}^{-}}$, and
\begin{equation}
\kappa_{+}^{2}(s) = \frac{k_{0}^{2}}{q_{0}} \left[ q_{0} \;+\frac{}{} p^{+}\left({\sf e}_{1}^{+} - {\sf e}_{2}^{+}\right) {\rm sd}^{2}\left(u^{+}\,|\,p^{\prime +}\right) \right],
\label{eq:kappa2_JII}
\end{equation}
where $\lambda > \lambda_{\Delta}$ and $u^{+} \equiv \xi\sqrt{{\sf e}_{1}^{+} - {\sf e}_{3}^{+}}$. We note that these two solutions do not hold simultaneously in parameter space $(\lambda,\nu)$.

Lastly, we have so far derived solutions of the squared-curvature equation \eqref{eq:K2_prime} expressed either in terms of the Weierstrass elliptic solution \eqref{eq:kappa_PWJ}, or the Jacobi elliptic solutions \eqref{eq:kappa2_JI}-\eqref{eq:kappa2_JII}. We note that, beside the initial-condition parameter $k_{0} = \kappa(0)$, these solutions depend on three parameters: the curvature parameters $(\lambda,\nu)$ and the scale parameter $q_{0}$. 

\section{\label{sec:LS_J}Langer-Singer Parametrization of the Curvature Solution}
 
\begin{figure}
\epsfysize=2in
\epsfbox{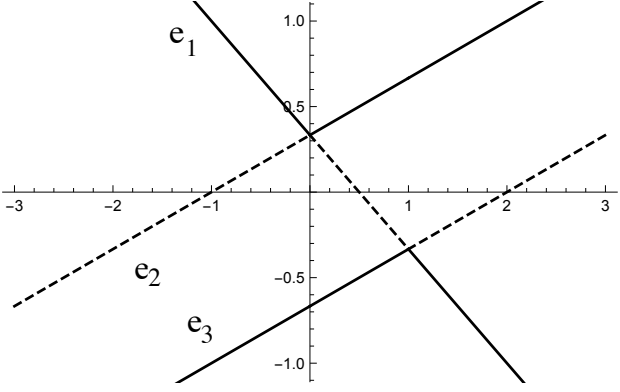}
\caption{Plots of the cubic roots ${\sf e}_{3}(m) \leq {\sf e}_{2}(m) \leq {\sf e}_{1}(m)$ as functions of $m$ in the range $-3 \leq m \leq 3$. The classical Langer-Singer range is $0 \leq m \leq 1$, where ${\sf e}_{1} = 
{\sf e}_{2}$ at $m = 0$ and ${\sf e}_{2} = {\sf e}_{3}$ at $m = 1$.}
\label{fig:root_LS}
\end{figure}

 The classical parametrization for the Jacobi elliptic solution of the squared-curvature equation \eqref{eq:K2_prime} was introduced by Langer and Singer \cite{Langer_Singer_1984}, with the Jacobi modulus $0 \leq m \leq 1$ chosen as
 \begin{equation}
  m \;\equiv\; ({\sf e}_{1} -  {\sf e}_{2})/({\sf e}_{1} - {\sf e}_{3}) \;=\; {\sf e}_{1} -  {\sf e}_{2},
 \end{equation}
with $ {\sf e}_{1} - {\sf e}_{3} \equiv 1$, so that $\lambda$ and $\nu$ become functions of $(m,q_{0})$. With these choices, we obtain the 
 two relations ${\sf e}_{1} = 1 + {\sf e}_{3} = m + {\sf e}_{2}$, from which, using the identity ${\sf e}_{2} = -\,{\sf e}_{1} - {\sf e}_{3}$, we find the three cubic roots:
\begin{equation}
\left. \begin{array}{l}
{\sf e}_{1}(m) = (1+m)/3 \\
{\sf e}_{2}(m) = (1 - 2\,m)/3 \\
{\sf e}_{3}(m) = (m-2)/3
\end{array} \right\}
\label{eq:classical_roots}
\end{equation}
in the  classical range $0 \leq m \leq 1$. These roots are shown in Fig.~\ref{fig:root_LS}, where ${\sf e}_{1}$ and ${\sf e}_{3}$ are shown as the piecewise continuous top and bottom solid lines, respectively, while 
${\sf e}_{2}$ is shown as the piecewise continuous dashed line. According to Fig.~\ref{fig:root_LS}, the cubic roots can be extended outside the classical range $0 \leq m \leq 1$ as follows. In the range $m \leq 0$, we find
\begin{equation}
\left. \begin{array}{l}
{\sf e}_{1}(m) = (1 - 2\,m)/3 \\
{\sf e}_{2}(m) = (1 + m)/3 \\
{\sf e}_{3}(m) = (m-2)/3
\end{array} \right\},
\label{eq:m_neg}
\end{equation}
while in the range $m \geq 1$, we find
\begin{equation}
\left. \begin{array}{l}
{\sf e}_{1}(m) = (1 +m)/3 \\
{\sf e}_{2}(m) = (m - 2)/3 \\
{\sf e}_{3}(m) = (1 - 2\,m)/3
\end{array} \right\},
\label{eq:m_plus}
\end{equation}
where the ranges are selected in order to preserve the ordering ${\sf e}_{3} \leq {\sf e}_{2} \leq {\sf e}_{1}$ for all values of $m$. 

\begin{figure}
\epsfysize=1.9in
\epsfbox{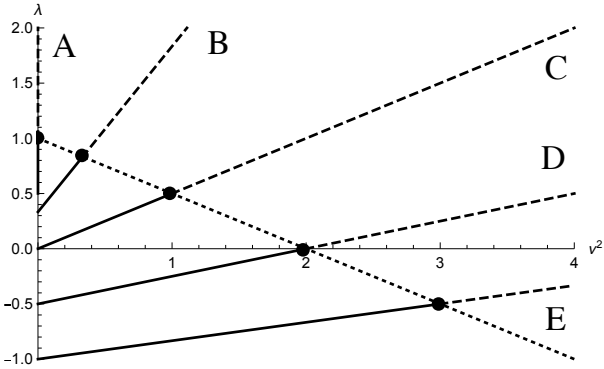}
\caption{Plots of $\lambda(m,q_{0})$ versus $\nu^{2}(m,q_{0})$ in the classical range $0 \leq m \leq q_{0} \leq 1$ (solid) and the extended range $m \leq 0$ (dashed) for (A) $q_{0} = 1$, (B) $q_{0} = 3/4$, (C) $q_{0} = 1/2$, (D) $q_{0} = 1/3$, and (E) $q_{0} = 1/6$. The dots on the dotted line, corresponding to the boundary $\lambda(0,q_{0}) = 1 - \nu^{2}(0,q_{0})/2$ at $m = 0$, show the transition from the classical range to the extended range.}
\label{fig:Classical}
\end{figure}

Returning to Eqs.~\eqref{eq:cubic_a} and \eqref{eq:classical_roots}-\eqref{eq:m_neg}, now written as $\lambda = \frac{3}{2}\,(1 - {\sf e}_{a}/q_{0})$, we find the curvature function
\begin{equation}
\lambda(m,q_{0}) \;=\; \frac{3}{2} \;-\; \frac{(1+m)}{2\,q_{0}} 
\label{eq:LS_lambda} 
\end{equation}
for all values of $(m,q_{0})$ satisfying the condition $m \leq q_{0} \leq 1$. For the torsion function $\nu^{2}(m,q_{0})$, we use Eq.~\eqref{eq:nu2_def} and find
\begin{equation}
\nu^{2} = (1- q_{0})\,(q_{0} - m)/q_{0}^{2},
\label{eq:LS_torsion} 
\end{equation}
which is positive in the range $m \leq q_{0} \leq 1$. The plots of $\lambda(m,q_{0})$ versus $\nu^{2}(m,q_{0})$ in the classical range $0 \leq m \leq q_{0} \leq 1$ (solid) and the extended range $m \leq 0$ (dashed) are shown in Fig.~\ref{fig:Classical}. Each straight line in Fig.~\ref{fig:Classical} is parametrized by $m \leq q_{0}$ for a fixed value of $q_{0} \leq 1$. The torsionless case ($\nu = 0$) is divided into two segments on the $\lambda$-axis: the upper segment $\lambda = 1 - m/2 \geq 1/2$ for $q_{0} = 1$ and $m \leq 1$; and the lower segment $\lambda = 1 - 1/(2m) \leq 1/2$ for $0 \leq q_{0} = m \leq 1$. In what follows, only the ranges $m \leq 0$ and $0 \leq m \leq q_{0} \leq 1$ will be explored since they entirely cover the parameter space 
$(\nu^{2},\lambda)$ in Fig.~\ref{fig:Classical}.

The Weiertrass invariant functions \eqref{eq:g23_Delta} are 
\begin{equation}
\left. \begin{array}{l}
g_{2}(m) = (4/3)\,(m^{2} - m + 1) \\
g_{3}(m) = (4/27)\,(m+1) (1-2m) (m-2) \\
\Delta(m) = 16\,m^{2} (1 - m)^{2}
\end{array} \right\}. 
\end{equation}
Here, we note that $g_{2}(m) > 0$ for all real values of $m$, while $\Delta(m) > 0$ is positive for all values of $m$ except when $m = 0$ or $m = 1$, where it vanishes. The function $g_{3}(m)$ is positive either when $m < -1$ or $\frac{1}{2} < m < 2$, while it is negative either for $-1 < m < \frac{1}{2}$ or $m > 2$. In the classical Langer-Singer range $0 \leq m \leq 1$, $g_{3}(m)$ therefore changes sign only once.

 \begin{figure}
\epsfysize=1.8in
\epsfbox{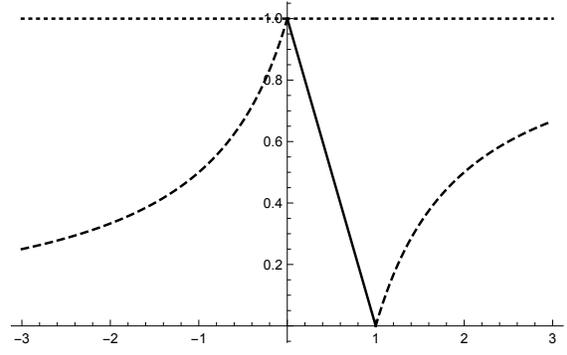}
\caption{Plot of the Jacobi modulus $0 \leq p(m) \leq 1$ as a function of $m$ in the range $-3 \leq m \leq 3$. The classical range $0 \leq m \leq 1$ is represented as a solid line while the extended ranges $m \leq 0$ and $m \geq 1$ are represented as dashed curves.}
\label{fig:parameter_m}
\end{figure}

\begin{figure}
\epsfysize=2in
\epsfbox{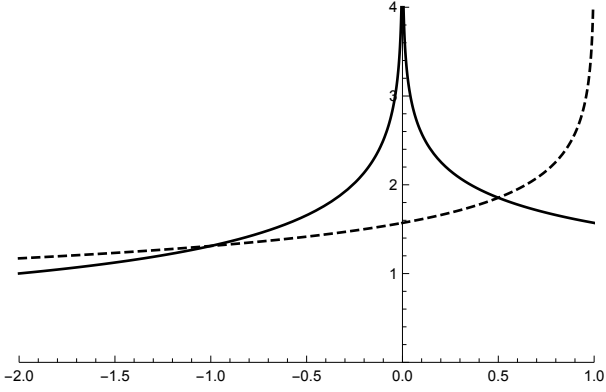}
\caption{Plots of the real half-period $\omega_{1}(m)$ (solid) and the imaginary half-period $|\omega_{3}|(m)$ (dashed) of the Weierstrass elliptic function as functions of $m$ in the range $-2 \leq m \leq 1$.}
\label{fig:period_knot}
\end{figure}

Using the cubic-root parameterization \eqref{eq:classical_roots}-\eqref{eq:m_plus}, the Jacobi modulus is now defined as
\begin{equation}
p(m) = \frac{{\sf e}_{2} - {\sf e}_{3}}{{\sf e}_{1} - {\sf e}_{3}}  = \left\{ \begin{array}{lr}
 1/m^{\prime}  &  (m \leq 0) \\
m^{\prime} \equiv 1 - m  &  (0 \leq m \leq 1) \\
1 - 1/m & (m \geq 1)
 \end{array} \right.
 \label{eq:p_m}
 \end{equation}
 and
 \begin{equation}
p^{\prime}(m) = \frac{{\sf e}_{1} - {\sf e}_{2}}{{\sf e}_{1} - {\sf e}_{3}}  = \left\{ \begin{array}{lr}
-m/m^{\prime} &  (m \leq 0) \\
m  &  (0 \leq m \leq 1) \\
1/m & (m \geq 1)
 \end{array} \right.
 \label{eq:pprime_m}
 \end{equation}
with 
\begin{equation}
{\sf e}_{1} - {\sf e}_{3} =  \left\{ \begin{array}{lr}
m^{\prime}  &  (m \leq 0) \\
1   &  (0 \leq m \leq 1) \\
m & (m \geq 1)
 \end{array} \right.
 \label{eq:e13_m}
 \end{equation}
 From Fig.~\ref{fig:parameter_m}, we note that, for all values of $m$, the Jacobi moduli \eqref{eq:p_m}-\eqref{eq:pprime_m} satisfy the relations $0 \leq p(m) \leq 1$ and $0 \leq p^{\prime}(m) \equiv 1 - p(m) \leq 1$, where $p(0) = 1$ and $p^{\prime}(1) = 1$, when the cubic roots are equal:  ${\sf e}_{2} = {\sf e}_{1}$ and ${\sf e}_{2} = {\sf e}_{3}$, respectively. The half-periods are, therefore, expressed in terms of the complete elliptic integrals
\begin{eqnarray}
 i\,{\sf K}(m) & = & i \left\{ \begin{array}{lr}
{\sf K}(-\,m/m^{\prime})/\sqrt{m^{\prime}} & (m \leq 0) \\
{\sf K}(m)  & (0 \leq m \leq 1) 
\end{array} \right. \nonumber \\
 & \equiv & \omega_{3}(m) 
\label{eq:omega3_m}
\end{eqnarray}
where we used the relation \eqref{eq:K_neg}, and 
\begin{eqnarray}
{\sf K}(m^{\prime}) & = & \left\{ \begin{array}{lr}
{\sf K}(1/m^{\prime})/\sqrt{m^{\prime}} - i\,{\sf K}(m) & (m \leq 0) \\
{\sf K}(m^{\prime})  & (0 \leq m \leq 1) 
\end{array} \right. \nonumber \\
& \equiv & \left\{ \begin{array}{lr}
\omega_{1}(m) - \omega_{3}(m) & (m \leq 0) \\
\omega_{1}(m) & (0 \leq m \leq 1) 
\end{array} \right.
\label{eq:omega1_m}
\end{eqnarray}
where we used the relation \eqref{eq:K_plus}. These half-period functions are shown in Fig.~\ref{fig:period_knot}, where $\omega_{1}(m)$ (solid) and $\omega_{3}(m)$ (dashed) become infinite at $m = 0$ 
$({\sf e}_{1} = {\sf e}_{2})$ and $m = 1$ $({\sf e}_{2} = {\sf e}_{3})$, respectively, where $\Delta(m)$ vanishes. 

\subsection{Jacobi parameter space}

\begin{figure}
\epsfysize=1.8in
\epsfbox{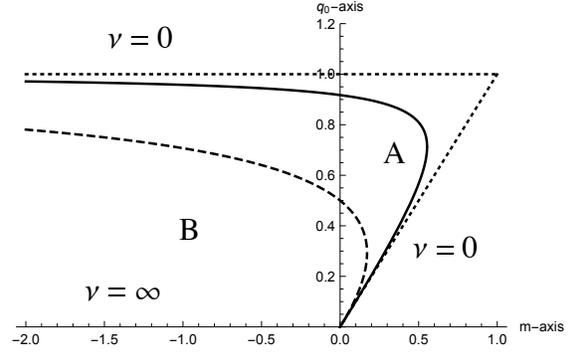}
\caption{Jacobi parameter space $(m,q_{0})$ defined by Eq.~\eqref{eq:q0_nu_I}, corresponding to the condition $\nu^{2} \geq 0$. The dotted lines correspond to the torsionless $(\nu = 0)$ case, while the solid and dashed curves correspond to $\nu = 0.3$ and $\nu = 1$, respectively.}
\label{fig:parameter_knot}
\end{figure}

According to Eq.~\eqref{eq:LS_torsion}, the initial torsion $\tau_{0} \equiv \tau(0) \neq 0$ is real and non-vanishing provided the parameters $(m,q_{0})$ satisfy $m < q_{0} < 1$, while $\tau_{0}$ vanishes for $q_{0} = 1$ or $q_{0} = m < 1$. Hence, using the parametrization \eqref{eq:LS_torsion} with a fixed value of $\nu^{2}$, the parameter space $(m < 1,q_{0})$ is defined by the boundaries
\begin{equation}
q_{0}^{\pm}(m,\nu) = \frac{(1+m) \pm \sqrt{(1 - m)^{2} - 4\,\nu^{2}\,m}}{2\;(1 + \nu^{2})},
\label{eq:q0_nu_I}
\end{equation}
where $\pm$ define the upper $(+)$ and lower $(-)$ boundaries in Fig.~\ref{fig:parameter_knot}. Here, we note that region A ($0 \leq m \leq q_{0} \leq 1)$ represents the classical parameter space explored by Langer and Singer \cite{Langer_Singer_1984}, which is shown as a dotted triangle with boundaries at $q_{0}^{+}(m,0) = 1$ and $Q_{0}(m,0) = m$. The region B ($m < 0 \leq q_{0} \leq 1$) of the parameter space remained unexplored until the present work. 

Secondly, we note that as the dimensionless torsion parameter $\nu$ increases, the region A decreases. In the limit $\nu \rightarrow \infty$, the region A disappears completely and the parameter space shrinks to $m \leq 0$ (in region B) on the line $q_{0} = 0$ (see Fig.~\ref{fig:parameter_knot}). The dotted lines, with $q_{0} = 1$ and $q_{0} = m$ (for $m < 1$), show the torsionless $(\nu = 0)$ boundaries of the parameter space $(m,q_{0})$. The $q_{0}$-axis (with $m = 0$) corresponds to the case $\lambda = \lambda_{\Delta}$ where $\Delta$ vanishes. The parameter spaces corresponding to $\nu = 0.3$ (solid) and $\nu = 1$ (dashed) are also shown, while the parameter space corresponding to the large-torsion limit $(\nu \gg 1)$ is the line $m < 0$ with $q_{0} = 0$. 

\begin{figure}
\epsfysize=1.5in
\epsfbox{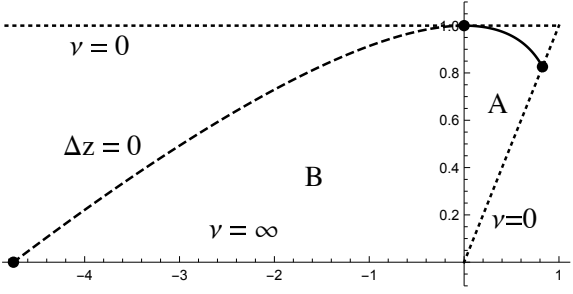}
\caption{Plot of $Q_{0}(m)$ (obtained from the periodicity constraint $\Delta z = 0$) as a function of $m$ in the classical range $0 \leq m \leq m_{0}^{-} = 0.82611...$ (solid curve) and the extended range 
$m_{0}^{+} = -4.75076... \leq m \leq 0$ (dashed curve), drawn within the Langer-Singer parameter space $(m,q_{0})$, with boundaries at $q_{0} = 1$ and $q_{0} = m$, where the torsion parameter $\nu$ is zero.}
\label{fig:Q0_J}
\end{figure}

We will see in Sec.~\ref{sec:elastica} that the elastica knots considered in this work have a natural cylindrical geometry, with both the cylindrical radius $\rho(s)$ and the vertical position $z(s)$ required to be periodic functions of $s$, while the azimuthal angle $\theta(s)$ is required to satisfy certain conditions in order for the elastica knot to be closed. In particular, we will see that the periodicity condition $\Delta z = z(s+S) - z(s) = 0$ requires that the scale factor $q_{0}$ be treated as a function of the Langer-Singer modulus $m$:
\begin{equation}
q_{0} \;=\; Q_{0}(m) \;\equiv\; 2\,\frac{{\sf E}(m)}{{\sf K}(m)} \;-\; (1 - m),
\label{eq:q_LS}
\end{equation}
where ${\sf E}(m)$ is the complete elliptic integral of the second kind. As shown in Fig.~\ref{fig:Q0_J}, we note that $Q_{0}(m_{0}^{-}) = m_{0}^{-}$ (i.e., $\nu = 0$), when $m = m_{0}^{-} = 0.82611...$, where $2\,{\sf E}(m_{0}^{-}) = {\sf K}(m_{0}^{-})$. We also note that $Q_{0}(0) = 1$ (i.e., $\nu = 0$), and $Q_{0}(m_{0}^{+}) = 0$, when $m_{0}^{+} = -\,m_{0}^{-}/(1 - m_{0}^{-}) = -4.75076..$, with $\nu \rightarrow \infty$ as $m \rightarrow m_{0}^{+}$.

\begin{figure}
\epsfysize=1.9in
\epsfbox{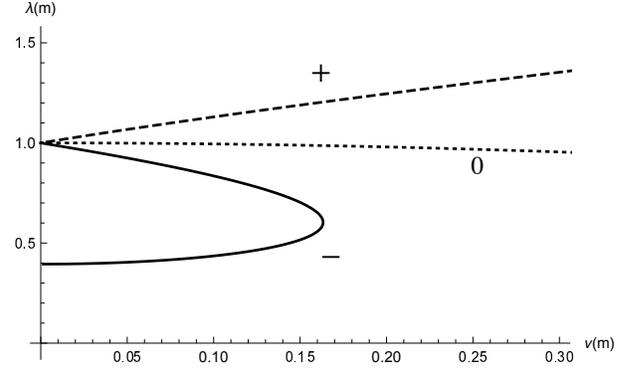}
\caption{Parametric plot of $\lambda(m)$ versus $\nu(m)$ as functions of $m$ in the classical range $0 \leq m \leq m_{0}^{-}$ (solid curve), where $(\nu,\lambda) = (0,1)$ at $m = 0$ and $(\nu,\lambda) =(0,1-1/2m)$ at $m = m_{0}^{-}$. In the extended range $m < 0$ (dashed curve), the point $(\nu,\lambda) \rightarrow (\infty,\infty)$ as $m \rightarrow m_{0}^{+}$, since $Q_{0}(m) \rightarrow 0$. The dotted curve corresponds to $\lambda = \lambda_{\Delta}(\nu) = 1 - \nu^{2}/2$.}
\label{fig:nu_J}
\end{figure}

With Eq.~\eqref{eq:q_LS}, we can construct the one-parameter functions
\begin{eqnarray}
\nu(m) & \equiv &  \sqrt{\left(\frac{1}{Q_{0}(m)} -1 \right)\,\left( 1 -\frac{m}{Q_{0}(m)}\right)}, \label{eq:nu_m} \\
\lambda(m) & \equiv & \frac{3}{2} \;-\; \frac{m+1}{2\,Q_{0}(m)},
\end{eqnarray}
and generate the parametric plot shown in Fig.~\ref{fig:nu_J}. We see that the parametric curve $(\nu(m),\lambda(m))$ is bounded in the classical range $0 \leq m \leq m_{0}^{-}$ (solid curve) and reaches a maximum for $\nu(m^{*}) = 0.1632...$ at $m^{*} = 0.6455...$. In the extended range $m_{0}^{+} \leq m \leq 0$, the parametric curve $(\nu(m),\lambda(m))$ is unbounded and $(\nu,\lambda) \rightarrow (\infty,\infty)$ as $m \rightarrow m_{0}^{+}$.

 \subsection{Langer-Singer elliptic solutions}
          
We now return to Eqs.~\eqref{eq:kappa2_JI}-\eqref{eq:kappa2_JII} to express the curvature solution $\kappa^{2}(s,m,q_{0})$ as a function of Jacobi elliptic functions. For the classical case $\lambda < \lambda_{\Delta}$ (with $0 \leq m \leq 1$), the Jacobi elliptic solution \eqref{eq:kappa2_JI} yields the Langer-Singer solution \cite{Langer_Singer_1984}
 \begin{equation}
 \kappa^{2}(s,m,q_{0}) \;=\; k_{0}^{2} \left[ 1 - \frac{m}{q_{0}}\;{\rm sn}^{2}\left(\xi\,|\,m\right)\right].
 \label{eq:kappa_J_I}
 \end{equation} 
When this solution is evaluated at the mid-point $\xi = {\sf K}(m)$, we find
\begin{eqnarray*}
\kappa^{2}(S/2,m,q_{0}) & = & k_{0}^{2} \left[ 1 - \frac{m}{q_{0}}\;{\rm sn}^{2}({\sf K}(m)|m) \right] \\ 
 & = & k_{0}^{2}\;\left( 1 - \frac{m}{q_{0}}\right) \;<\; k_{0}^{2},
 \end{eqnarray*}
 where we used ${\rm sn}^{2}({\sf K}(m)|m) = 1$. Hence, since $\lambda < \lambda_{\Delta}$, the initial curvature $k_{0}$ is a maximum.
 
\begin{figure}
\epsfysize=1.8in
\epsfbox{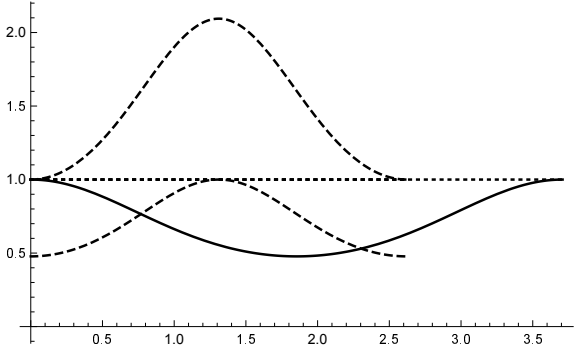}
\caption{Plot of Eq.~\eqref{eq:kappa_Jacobi} as a function of $\xi$ in the range $0 \leq \xi \leq 2\,{\sf K}(m)$, with the constraint $q_{0} = Q_{0}(m)$ defined in Eq.~\eqref{eq:q_LS}. Here, $m = 1/2$ (solid curve below dotted line) and $m = -1$ (dashed curve above dotted line) are shown since they are related by the symmetry $n(-1) = 1/2$. The dashed curve below the dotted line shows Eq.~\eqref{eq:kappa_J_DP}, which is identical to the upper dashed curve when it is divided by the factor $(1 + |m|/q_{0})$. }
\label{fig:K_knot}
\end{figure}
 
\begin{figure}
\epsfysize=1.8in
\epsfbox{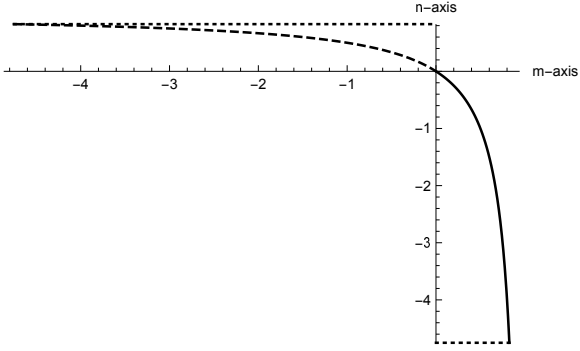}
\caption{Plot of $n(m) = -m/(1-m)$ versus $m$. The range $m_{0}^{+} < m < m_{0}^{-}$, where $m_{0}^{-} = n(m_{0}^{+}) = 0.82611... \equiv n_{0}^{+}$ and $m_{0}^{+} = n(m_{0}^{-}) = -\,4.75076...\equiv n_{0}^{-}$, is transformed into $n_{0}^{-} < n < n_{0}^{+}$.}
\label{fig:n_m}
\end{figure}

 For the extended case $\lambda > \lambda_{\Delta}$ (i.e., $m < 0$ ), on the other hand, the Jacobi elliptic solution \eqref{eq:kappa2_JII} yields
 \begin{equation}
 \kappa^{2}(s,m,q_{0}) = k_{0}^{2} \left[ 1 + \frac{n}{q_{0}}\,{\rm sd}^{2}\left(\xi/\sqrt{1-n}\left|n\right.\right) \right],
 \label{eq:kappa_J_II}
 \end{equation}  
 where the modulus transformation
 \begin{equation}
 n(m) \;=\; \frac{-\,m}{1-m} \;\leftrightarrow\; m(n) \;=\; \frac{-\,n}{1-n}
 \label{eq:n_m}
 \end{equation}
is used to relate a negative value of $m$ to a positive value $0 < n(m) < 1$. Figure \ref{fig:n_m} shows that this transformation introduces a symmetry between a solution for $m(n)$ and a solution for $n(m)$, e.g., the solution with $m = -1$ is related by symmetry to the solution with $m = 1/2$ since $n(-1) = 1/2$ and $n(1/2) = -1$. When Eq.~\eqref{eq:kappa_J_II} is evaluated at $\xi = {\sf K}(n)\sqrt{1 -n}$, we find
\begin{eqnarray*} 
\kappa^{2}(S/2,m,q_{0}) & = & \frac{k_{0}^{2}}{q_{0}} \left[ q_{0} + \frac{|m|\;{\rm sn}^{2}({\sf K})}{(1+|m|)\,{\rm dn}^{2}({\sf K})} \right] \\
 & = & k_{0}^{2}\;\left( 1 \;+\; \frac{|m|}{q_{0}}\right) \;>\; k_{0}^{2},
 \end{eqnarray*}
where we used ${\rm sn}^{2}({\sf K}) = 1$ and ${\rm dn}^{2}({\sf K}) = 1/(1+|m|)$. It can be shown that Eq.~\eqref{eq:kappa_J_II} can be written as
\begin{equation}
 \kappa^{2}(s,m,q_{0})/\wh{k}_{0}^{2} \;=\; 1 - \frac{n}{\wh{q}}\,{\rm sn}^{2}\left(\wh{\xi} - {\sf K}(n)\left|\;n\frac{}{}\right.\right),
 \label{eq:kappa_J_DP}
 \end{equation}
 where $\wh{\xi} = \xi/\sqrt{1-n(m)}$, $\wh{q} = q_{0} + n(m)\,(1- q_{0})$, and $\wh{k}_{0}^{2} = k_{0}^{2} (1 + |m|/q_{0})$. Hence, since $\lambda > \lambda_{\Delta}$, the initial curvature $k_{0}$ is a minimum.
 
If we use the transformation \eqref{eq:sn_transform}, we find that the solutions \eqref{eq:kappa_J_I}-\eqref{eq:kappa_J_DP} can be represented as
\begin{equation}
 \kappa^{2}(s,m,q_{0})/k_{0}^{2} \;=\; 1 \;-\frac{}{} \frac{m}{q_{0}}\;{\rm sn}^{2}\left(\xi\,|\,m\right),
 \label{eq:kappa_Jacobi}
 \end{equation} 
for all values $m \leq q_{0}$ (including $m \leq 0$). The plot of Eq.~\eqref{eq:kappa_Jacobi} is shown in Fig.~\ref{fig:K_knot}, with $q_{0} = Q_{0}(m)$ defined by Eq.~\eqref{eq:q_LS} in accordance to the periodicity constraint $\Delta z = 0$, with $m = 1/2$ (solid curve below dotted line) and $m = -1$ (dashed curve above dotted line). We note that for $0 < m \leq Q_{0}(m)$ (solid curve), the squared curvature \eqref{eq:kappa_Jacobi} is limited in range to $0 < \kappa^{2}(\xi,m)/k_{0}^{2} \leq 1$, while, for $m \leq 0$ (upper dashed curve), the squared curvature \eqref{eq:kappa_Jacobi} satisfies $\kappa^{2}(\xi,m)/k_{0}^{2} \geq 1$, with a maximum becoming infinite when $Q_{0}(m)$ vanishes. We also note that the period $2\,{\sf K}(m)$ is larger for $m > 0$ compared to $m < 0$ (see dashed curve in Fig.~\ref{fig:period_knot}).

\section{Curvature and Torsion Functionals}

In this Section, we use the Jacobi elliptic solution \eqref{eq:kappa_Jacobi}, with the constraint \eqref{eq:q_LS}, to evaluate curvature and torsion functionals.

\subsection{Curvature functional}

We now return to the curvature functional \eqref{eq:F_Lambda}, with the constraint $|{\bf r}^{\prime}| = 1$ now implemented. Inserting the squared-curvature solution
\eqref{eq:kappa_Jacobi}, we evaluate the normalized curvature functional
\begin{eqnarray}
 &  &\ov{\cal F}(m,q_{0}) = \frac{1}{2k_{0}\,\wh{\kappa}}\int_{0}^{S}\kappa^{2}(s)\;ds \nonumber \\
 & = & \frac{1}{\sqrt{q_{0}\wh{\kappa}^{2}}} \left[ \int_{0}^{2\,{\sf K}(m)}{\rm dn}^{2}(\xi |m)\,d\xi - (1 - q_{0})\;2\,{\sf K}(m) \right] \nonumber \\
 & = & \frac{2}{\sqrt{q_{0}\wh{\kappa}^{2}}} \left[ \frac{1}{2}\,{\sf Z}(2{\sf K}(m)\,|\,m) + {\sf E}(m) - (1 - q_{0})\,{\sf K}(m) \right] \nonumber \\
  & = & \frac{2}{\sqrt{q_{0}\wh{\kappa}^{2}}} \left[ {\sf E}(m) \;-\frac{}{} (1 - q_{0})\,{\sf K}(m)\right],
\label{eq:F_S}
  \end{eqnarray}
where the normalizing factor
\begin{equation}
\wh{\kappa}^{2} \;\equiv\; \left\{ \begin{array}{lr}
1 & (m > 0) \\
 & \\
 1 - m/q_{0} & (m < 0)
 \end{array} \right. 
 \label{eq:kappa_factor}
 \end{equation}
 guarantees that the curvature functional is normalized with respect to the maximum curvature according to Eqs.~\eqref{eq:kappa_J_I}-\eqref{eq:kappa_J_DP}, ${\sf E}(m)$ denotes the complete elliptic integral of the second kind, and we use the simplified notation ${\sf Z}(\xi|m)$ for the Jacobi zeta function \cite{NIST_Jacobi}
\begin{eqnarray} 
{\sf Z}(\xi|m) & \equiv & {\sf Z}\left(\arcsin[{\rm sn}(\xi|m)]\left.\frac{}{}\right|m\right) \nonumber \\
 & \equiv & \int_{0}^{\xi}{\rm dn}^{2}(u|m)\,du \;-\; \frac{{\sf E}(m)}{{\sf K}(m)}\;\xi, 
\label{eq:JacobiZ_def}
\end{eqnarray}
which is periodic, ${\sf Z}(\xi + 2\,{\sf K}|m) = {\sf Z}(\xi|m)$, and vanishes at $\xi = 2\,{\sf K}(m)$. 

 \begin{figure}
\epsfysize=2in
\epsfbox{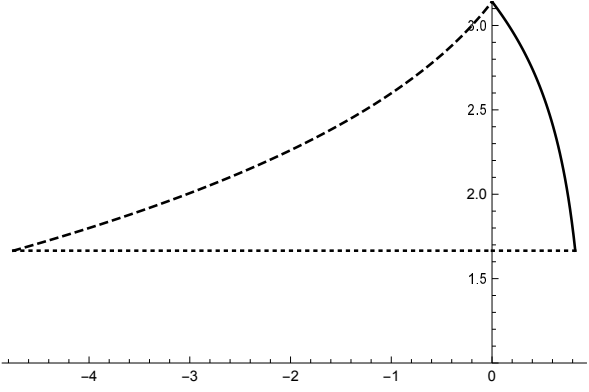}
\caption{Plot of the normalized curvature functional $\wh{\cal F}(m) \equiv \ov{\cal F}(m,Q_{0}(m))$ versus $m$ in the range $0 \leq m \leq m_{0}^{-}$ (solid curve) and $m_{0}^{+} \equiv n(m_{0}^{-}) \leq m \leq 0$ (dashed curve) for $q_{0} = Q_{0}(m)$. At the boundary $m = 0$ ($q_{0} = 1$), we find $\wh{\cal F}(0) = \pi$, while $\wh{\cal F}(m_{0}^{\pm}) = (2\,m_{0}^{-} - 1){\sf K}(m_{0}^{-})/\sqrt{m_{0}^{-}}$ (dotted horizontal line) at the end points $m = m_{0}^{\pm}$.}
\label{fig:curv_knot}
\end{figure}

Using the constraint $q_{0} = Q_{0}(m)$, Eq.~\eqref{eq:F_S} yields $\wh{\cal F}(m) \equiv \ov{\cal F}(m,Q_{0}(m))$, which is shown in Fig.~\ref{fig:curv_knot} (as a solid curve for $m > 0$ and a dashed curve for $m < 0$). At the boundary $m = 0$, we find $\wh{\cal F}(0) = \pi$. Lastly, we note that the normalized curvature functional satisfies the modulus symmetry
\begin{equation}
\wh{\cal F}(n(m)) \;=\; \ov{\cal F}(m),
\end{equation}
with $\wh{\cal F}(m_{0}^{\pm}) = (2 m_{0}^{-} - 1)\,{\sf K}(m_{0}^{-})/\sqrt{m_{0}^{-}}$. Hence, the curvature functional is numerically identical at $m = -1$ and $n(-1) = 1/2$.

\subsection{Torsion functionals}

We calculate the averaged (normalized) torsion
\begin{eqnarray}
\langle\tau\rangle(m,q_{0}) & \equiv & \frac{1}{k_{0}S\wh{\kappa}}\int_{0}^{S}\tau(s)\;ds \;=\; \frac{1}{S\,\wh{\kappa}}\int_{0}^{S}\frac{k_{0}\tau_{0}\,ds}{\kappa^{2}(s)} \nonumber \\
 & = & \frac{q_{0}\nu}{4\,\wh{\kappa}\omega_{3}}\int_{\omega_{a}}^{\omega_{a} + 2\,\omega_{3}}\frac{d\varphi}{\wp(\varphi) - \wp(\psi)},
\label{eq:avetor_def}
\end{eqnarray}
where, using Eq.~\eqref{eq:e1_chi}, $\psi$ is defined through $\wp(\psi) \equiv -\,2q_{0}\lambda/3 = {\sf e}_{a} - q_{0}$. Using the definition \eqref{eq:nu2_def}, we easily find that
$\wp^{\prime}(\psi) = 2\,q_{0}^{3/2}\,\nu$, so that the averaged torsion \eqref{eq:avetor_def} becomes
\begin{eqnarray}
\langle\tau\rangle(m,q_{0}) & = & \frac{1}{8\sqrt{q_{0}\wh{\kappa}^{2}}\,\omega_{3}}\int_{\omega_{a}}^{\omega_{a}+ 2\,\omega_{3}}\frac{\wp^{\prime}(\psi)\;d\varphi}{\wp(\varphi) - \wp(\psi)} \nonumber \\
 & = & \frac{1}{2\sqrt{q_{0}\wh{\kappa}^{2}}\,\omega_{3}} \left[ \omega_{3}\,\zeta(\psi) \;-\frac{}{} \psi\,\zeta(\omega_{3}) \right].
 \label{eq:tau_averaged}
\end{eqnarray} 
Next, we use the Langer-Singer parametrization, with the constraint $q_{0} \equiv Q_{0}(m)$, so that $\psi \equiv \psi(m)$ yields the special values 
$\psi(0) = \omega_{3}(0) = i\,\pi/2$ and $\psi(m_{0}^{\pm}) = -\,\omega_{2}(m_{0}^{\pm})$. 

\begin{figure}
\epsfysize=1.8in
\epsfbox{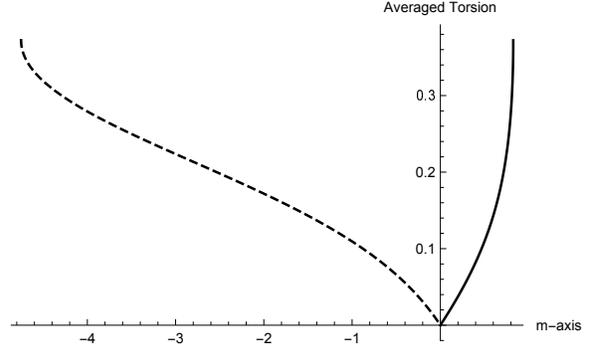}
\caption{Plot of the normalized averaged torsion $\langle\tau\rangle(m,Q_{0}(m))$ as a function of $m$ in the ranges $m_{0}^{+} < m \leq 0$ (dashed curve) and $0 \leq m \leq m_{0}^{-}$ (solid curve). The normalized averaged torsion has a finite value at $\langle\tau\rangle(m_{0}^{\pm}) = \pi/(4\sqrt{m_{0}^{-}}\,{\sf K}(m_{0}^{-}))$.}
\label{fig:ave_tor}
\end{figure}

Figure \ref{fig:ave_tor} shows the plot of the normalized averaged torsion $\langle\tau\rangle(m)$ as a function of $m$ in the ranges $m_{0}^{+} < m \leq 0$ (dashed curve) and $0 \leq m \leq m_{0}^{-}$ (solid curve). The averaged torsion has a finite value at $\langle\tau\rangle(m_{0}^{\pm}) = \pi/(4\sqrt{m_{0}^{-}}\,{\sf K}(m_{0}^{-}))$. At $m = 0$, we find $Q_{0}(0) = 1$, and the averaged torsion \eqref{eq:tau_averaged} is zero since $\omega_{3}\,\zeta(\omega_{3}) - \omega_{3}\,\zeta(\omega_{3}) = 0$ in the numerator of Eq.~\eqref{eq:tau_averaged}.

Another measure of the integrated torsion is the total torsion
\begin{eqnarray}
T(m) & \equiv & \frac{1}{2\pi}\int_{0}^{S}\tau(s)\;ds = \frac{1}{4\pi\,i}\int_{\omega_{a}}^{\omega_{a} + 2\,\omega_{3}}\frac{\wp^{\prime}(\psi)\;d\varphi}{\wp(\varphi) - \wp(\psi)} \nonumber \\
 & = & \frac{1}{\pi\,i} \left[ \omega_{3}\,\zeta(\psi) \;-\frac{}{} \psi\,\zeta(\omega_{3}) \right].
 \label{eq:T_m}
 \end{eqnarray}
At $m = 0$, we find $\psi(0) = \omega_{3}(0)$ so that $T(0) = 0$. At $m = m_{0}^{\pm}$, we find $\psi(m_{0}^{\pm}) = -\,\omega_{2}(m_{0}^{\pm})$ so that 
\[ T(m_{0}^{\pm}) \;=\; \frac{1}{\pi\,i} \left[ -\,\omega_{3}\,\zeta(\omega_{2}) \;+\frac{}{} \omega_{2}\,\zeta(\omega_{3}) \right] \;=\; \frac{1}{2}. \]

Lastly, we note that both torsion functionals \eqref{eq:tau_averaged} and \eqref{eq:T_m} satisfy the modulus symmetry 
\begin{equation}
\left. \begin{array}{rcl}
\langle\tau\rangle[n(m)] & = & \langle\tau\rangle(m) \\
 &  & \\
T[n(m)] & = & T(m)
\end{array} \right\}, 
\label{eq:torsion_nm}
\end{equation}
where $n(m) = -m/(1-m)$ and $q_{0} = Q_{0}(m)$. Hence, for example, these torsion functionals are numerically identical at $m = -1$ and $n(-1) = 1/2$.

\begin{figure}
\epsfysize=2in
\epsfbox{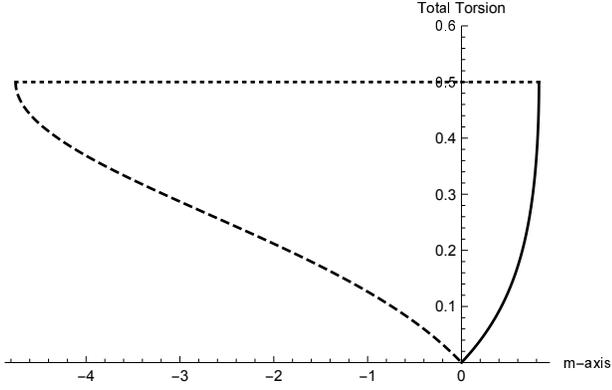}
\caption{Plot of the total torsion $T(m)$ as a function of $m$ in the ranges $m_{0}^{+} < m \leq 0$ (dashed curve) and $0 \leq m \leq m_{0}^{-}$ (solid curve). The dotted line is located at $T(m_{0}^{\pm}) = 1/2$.}
\label{fig:total_tor}
\end{figure}

\section{\label{sec:elastica}Elastica-Knot Space-Curve}

In this last Section, we will show that, with the help of the curvature $\kappa(s) \equiv |{\bf r}^{\prime\prime}|$, defined by the Weierstrass solution \eqref{eq:kappa_PW}, and the torsion constraint 
\begin{equation}
\tau(s) \;=\; \frac{\tau_{0}\;k_{0}^{2}}{\kappa^{2}(s)} \;\equiv\; \frac{\nu\;k_{0}^{3}}{2\;\kappa^{2}(s)},
\end{equation}
we will reconstruct the three-dimensional elastica-knot curve ${\bf r}(s)$.

\subsection{Cylindrical elastica-knot geometry}

The geometry of the elastica knots is determined by the fact that the vector ${\bf W}$ is constant, which suggests a cylindrical geometry. Hence, we begin with a cylindrical representation of the elastica-knot space curve:
\begin{equation}
{\bf r}(s) \;=\; \rho(s)\,\wh{\rho}(s) \;+\; z(s)\;\wh{\sf z},
\label{eq:cyl}
\end{equation}
where the constant unit vector 
\begin{equation}
\wh{\sf z} \;\equiv\; \frac{{\bf W}}{|{\bf W}|} \;=\; \alpha(s)\,\wh{\sf t} \;+\; \beta(s)\,\wh{\sf n} \;+\; \gamma(s)\,\wh{\sf b}
\label{eq:z_def}
\end{equation}
 is defined in terms of the constant vector \eqref{eq:W_const}, and the two unit vectors $\wh{\rho}(s)$ and $\wh{\theta}(s) \equiv \wh{\sf z}\btimes\wh{\rho}(s)$ are perpendicular to $\wh{\sf z}$ and may change their orientations as functions of $s$. In Eq.~\eqref{eq:z_def}, we have also defined the periodic functions
  \begin{equation}
 \left. \begin{array}{rcl}
\alpha(s) & = & {\cal R}^{2}\left(\kappa^{2} \;-\frac{}{}\lambda\,k_{0}^{2} \right)/2   \\
  &  & \\
\beta(s)  & = &  {\cal R}^{2}\kappa^{\prime}  \\
  &  & \\
\gamma(s) & = & {\cal R}^{2}k_{0}^{2}\tau_{0}/\kappa \;\equiv\; \mu\,k_{0}/\kappa
 \end{array} \right\},
 \label{eq:abc}
 \end{equation}
which satisfy $\alpha^{2} + \beta^{2} + \gamma^{2} \equiv 1$, and the magnitude of the constant vector
\begin{equation}
 |{\bf W}|^{2} \;=\; \frac{k_{0}^{4}}{4}\left[ (1 - \lambda)^{2} \;+\frac{}{} \nu^{2} \right] \;\equiv\; {\cal R}^{-4}
 \label{eq:R_def}
 \end{equation}
 introduces a useful length scale ${\cal R}(\lambda,\nu,k_{0})$. In Eq.~\eqref{eq:abc}, we also introduced a new dimensionless parameter $\mu$, defined as
 \begin{equation}
 \mu^{2}(\lambda,\nu) \;\equiv\; {\cal R}^{4}k_{0}^{2}\tau_{0}^{2} \;=\; \frac{\nu^{2}}{(1 - \lambda)^{2} + \nu^{2}} \;\leq\; 1.
 \label{eq:mu_def}
 \end{equation}

In order to construct the cylindrical unit vectors $(\wh{\rho},\wh{\theta})$, we need to construct a vector ${\bf W}_{\bot}$ that is perpendicular to ${\bf W}$. Here, we note that, since the vector ${\bf W}$ has a constant projection along the vector $\kappa\,\wh{\sf b}$: 
\begin{equation}
{\bf W}\bdot\kappa\wh{\sf b} \;=\; \kappa\,|{\bf W}|\;\wh{\sf z}\bdot{\sf b} \;=\; \kappa^{2}\tau \;\equiv\; k_{0}^{2}\tau_{0}, 
\end{equation}
which follows from the torsion constraint \eqref{eq:tau_eq}, we are free to define ${\bf W}_{\bot}$ as
\begin{equation}
{\cal R}^{2}{\bf W}_{\bot} \;\equiv\; \wh{\sf z} - \gamma^{-1}\;\bhat,
\end{equation} 
which immediately leads to the definitions
\begin{eqnarray}
\wh{\theta} & \equiv & \frac{{\bf W}_{\bot}}{|{\bf W}_{\bot}|} \;=\; \frac{\gamma\,\wh{\sf z} \;-\; \wh{\sf b}}{\sqrt{1 - \gamma^{2}}},  \label{eq:theta_def} \\
 \wh{\rho} & \equiv & \frac{{\bf W}\btimes\wh{\sf b}}{|{\bf W}\btimes\wh{\sf b}|} \;=\; \frac{\wh{\sf z}\btimes\wh{\sf b}}{\sqrt{1 - \gamma^{2}}}
 \label{eq:rho_def}
 \end{eqnarray}
 where
 \begin{equation}
{\cal R}^{2}\,|{\bf W}\btimes\wh{\sf b}| \;=\; \gamma\,{\cal R}^{2}|{\bf W}_{\bot}| \;\equiv\; \sqrt{1 - \gamma^{2}}.
 \end{equation}
 Here, we see that, as expected, we find
 \begin{eqnarray*}
 \wh{\rho}\btimes\wh{\theta} & = & \frac{(\wh{\sf z}\btimes\wh{\sf b})\btimes(\gamma\,\wh{\sf z} \;-\; \wh{\sf b})}{1 - \gamma^{2}} \\
  & = & \frac{\gamma\,\wh{\sf b} - \gamma^{2}\,\wh{\sf z} - \gamma\,\wh{\sf b} + \wh{\sf z}}{1 - \gamma^{2}} \;=\; \wh{\sf z}.
 \end{eqnarray*}
 
 With the unit vectors \eqref{eq:z_def} and \eqref{eq:theta_def}-\eqref{eq:rho_def}, we can now write the tangent unit vector as
 \begin{equation}
 \wh{\sf t} \;=\; {\bf r}^{\prime}(s) \;=\; \rho^{\prime}(s)\;\wh{\rho} \;+\; \rho(s)\theta^{\prime}(s)\;\wh{\theta} \;+\; z^{\prime}(s)\;\wh{\sf z},
 \label{eq:r_cyl}
 \end{equation}
 which yields the differential equations for the cylindrical coordinates $(\rho,\theta,z)$:
 \begin{equation}
 \left. \begin{array}{rcl}
 \rho^{\prime}(s) & = & \wh{\sf t}\bdot\wh{\rho} = \beta/\sqrt{1 - \gamma^{2}} \\
  && \\
 \rho(s)\,\theta^{\prime}(s) & = & \wh{\sf t}\bdot\wh{\theta} =  \alpha\,\gamma/\sqrt{1 - \gamma^{2}} \\
 && \\
  z^{\prime}(s) & = & \wh{\sf t}\bdot\wh{\sf z} \;=\; \alpha
 \end{array} \right\}.
 \label{eq:rtz_prime}
 \end{equation}
 Hence, once the functions $\rho(s)$ and $z(s)$ are solved as functions of $s$, then the azimuthal angle $\theta(s)$ can also be found, so that the three-dimensional space curve
 \begin{equation}
 {\bf r}(s) \;\equiv\; \wh{\sf z}\btimes\left(\frac{\rho(s)\;\wh{\sf b}(s)}{\sqrt{1 - \gamma^{2}(s)}}\right) \;+\; z(s)\;\wh{\sf z}
 \label{eq:3d_curve}
 \end{equation}
 is completely determined from the initial conditions.
 
 \subsection{Cylindrical solutions}
 
\begin{figure}
\epsfysize=2in
\epsfbox{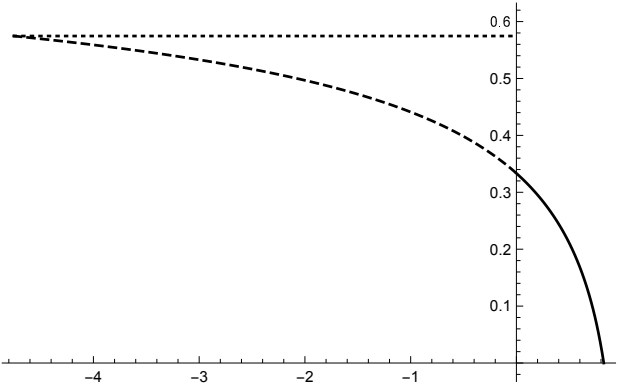}
\caption{Plot of $\mu^{2}(m,Q_{0}(m))$ as a function of $m$ in the ranges $m_{0}^{+} < m \leq 0$ (dashed curve) and $0 \leq m \leq m_{0}^{-}$ (solid curve). At the boundary $m = 0$, we find $\mu^{2}(0,1) = 1/3.$}
\label{fig:mu}
\end{figure}

From the differential equations \eqref{eq:rtz_prime}, we now obtain explicit solutions for the radial coordinate $\rho(s)$, the vertical coordinate $z(s)$, and the azimuthal angle $\theta(s)$. Using the Langer-Singer parametrization, Eq.~\eqref{eq:mu_def} is now expressed as
 \begin{equation}
 \mu^{2}(m,q_{0}) \;=\; \frac{4\,(1-q_{0})\,(q_{0} - m)}{(1+m - q_{0})^{2} + 4\,(1-q_{0})\,(q_{0} - m)},
 \end{equation}
 which is shown in Fig.~\ref{fig:mu}. We note that $\mu^{2}(m,Q_{0}(m))$ vanishes at $m = m_{0}^{-}$ and it reaches a maximum value $4\,|m_{0}^{+}|/(1+|m_{0}^{+}|)^{2} < 1$ at $m = m_{0}^{+}$.
 
\begin{figure}
\epsfysize=2in
\epsfbox{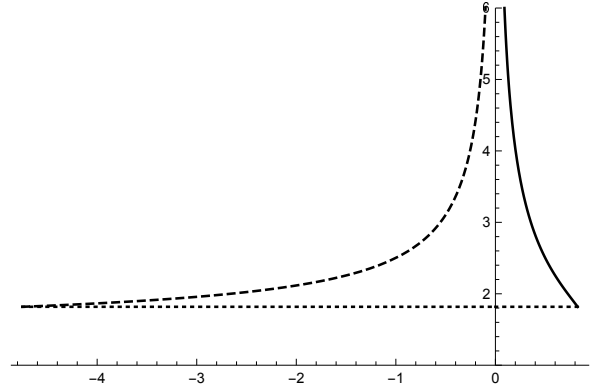}
\caption{Plot of the normalized radius $\wh{\cal R}(m)$ as a function of $m$ in the ranges $m_{0}^{+} < m \leq 0$ (dashed curve) and $0 \leq m \leq m_{0}^{-}$ (solid curve).}
\label{fig:R}
\end{figure}

Figure \ref{fig:R} shows a plot of the normalized radius
\begin{equation}
\wh{\cal R}(m) \;\equiv\; k_{0}\wh{\kappa}(m)\;{\cal R}(m,Q_{0}(m)),
\label{eq:R_m}
\end{equation}
where $\wh{\kappa}(m)$ is the normalization factor \eqref{eq:kappa_factor}. We note that Eq.~\eqref{eq:R_m} is infinite at $m = 0$, while it is finite at $m = m_{0}^{\pm}$: $\wh{\cal R}(m_{0}^{\pm}) = 2\sqrt{m_{0}^{-}}$. We also note that the normalized radius \eqref{eq:kappa_factor} satisfies the modulus symmetry $\wh{\cal R}(n(m)) = \wh{\cal R}(m)$.

 \subsubsection{Radial solution}
 
 The solution for the cylindrical radius $\rho(s)$ is easily obtained from 
 \[ \rho^{\prime} = \frac{{\cal R}^{2}\;\kappa\,\kappa^{\prime}}{\sqrt{\kappa^{2} - \mu^{2}\,k_{0}^{2}}} \equiv \frac{d}{ds}\left({\cal R}^{2}
 \sqrt{\kappa^{2} - \mu^{2}\,k_{0}^{2}}\right). \]
Using the initial condition 
\begin{equation}
\rho(0) \;=\; k_{0}{\cal R}^{2}\,\sqrt{1 - \mu^{2}}, 
\label{eq:rho_0}
\end{equation}
which, according to Fig.~\ref{fig:mu}, does not vanish, we find the periodic solution
 \begin{eqnarray}
 \rho(s) & = & {\cal R}^{2}\;\sqrt{\kappa^{2}(s) - \mu^{2}\,k_{0}^{2}} \nonumber \\
 & = & k_{0}{\cal R}^{2} \sqrt{\left(1 -\mu^{2}\right) \;-\; \frac{m}{q_{0}}\;{\rm sn}^{2}(\xi|m)}.
 \label{eq:rho_sol}
 \end{eqnarray}
Hence, the cylindrical radius evaluated at the midpoint $s = S/2$ is expressed as
 \begin{equation}
\rho(S/2) \;=\; k_{0}{\cal R}^{2}\;\sqrt{1 - \mu^{2} - m/q_{0}},
\label{eq:rho_mid}
\end{equation} 
which is $\rho(S/2) < \rho(0)$ for $0 < m < m_{0}^{-}$ and $\rho(S/2) > \rho(0)$ for $m_{0}^{+} < m < 0$. At $m = m_{0}^{-}$, we find that $\rho(S/2)$ vanishes, since $\mu^{2}$ vanishes and $Q_{0}(m_{0}^{-}) = m_{0}^{-}$, while at $m = m_{0}^{+}$, $\rho(S/2)$ becomes infinite, since $Q_{0}(m_{0}^{+}) = 0$.
 
Lastly, we note that the radial solution \eqref{eq:rho_sol} implies that the three-dimensional curve \eqref{eq:3d_curve} can also be expressed as
\begin{equation}
{\bf r}(s) \;=\; {\cal R}^{2}\,\kappa(s)\;\wh{\sf z}\btimes\wh{\sf b}(s) \;+\; z(s)\;\wh{\sf z},
\label{eq:3d_curve_kappa}
\end{equation}
where the orientation of the binormal unit vector $\wh{\sf b}(s)$ changes as a result of torsion $\tau(s)$: $\wh{\sf b}^{\prime} \equiv -\,\tau\,\wh{\sf n}$.

 \subsubsection{Vertical solution}
 
The solution for the vertical position $z(s)$ is also easily obtained from the equation $z^{\prime} = \frac{1}{2}\,{\cal R}^{2}\;(\kappa^{2} - \lambda\,k_{0}^{2})$. Assuming that $z(0) = 0$, we use the Langer-Singer parametrization to obtain
 \begin{eqnarray}
 z(s) & = & \frac{1}{2}\,{\cal R}^{2}\; \left( \int_{0}^{s}\;\kappa^{2}(s)\,ds  \;-\; \lambda\;k_{0}^{2}\,s \right) \nonumber \\
  & = &  \frac{k_{0}{\cal R}^{2}}{\sqrt{q_{0}}}\;{\sf Z}(\xi|m) \nonumber \\
 &  &+\; \frac{k_{0}{\cal R}^{2}\,\xi}{2\sqrt{q_{0}}}\;\left( 2\,\frac{{\sf E}(m)}{{\sf K}(m)} \;-\; (1 + q_{0} - m) \right),
 \end{eqnarray}
which is a periodic function of $\xi \equiv k_{0}s/2\sqrt{q_{0}}$
\begin{equation}
z(\xi) \;\equiv\; \frac{k_{0}{\cal R}^{2}}{\sqrt{q_{0}}}\,{\sf Z}(\xi|m),
\label{eq:z_minus_LS}
\end{equation}
with the Jacobi zeta function defined in Eq.~\eqref{eq:JacobiZ_def}, only if $q_{0}$ satisfies the constraint $\Delta z(m,q_{0}) \equiv 0$, where
\begin{eqnarray*} 
\Delta z(m,q_{0}) & \equiv & z(s + S) - z(s) \\
 & = & \frac{k_{0}^{2}{\cal R}^{2}S}{2\,q_{0}}\left[ 2\,\frac{{\sf E}(m)}{{\sf K}(m)} \;-\; (1 - m) \;-\; q_{0} \right],
 \end{eqnarray*}
which yields Eq.~\eqref{eq:q_LS}.

 \subsubsection{Azimuthal-angle solution}
 
 Lastly, the solution for the azimuthal angle $\theta(s)$ is obtained from
 \[ \rho\;\theta^{\prime} \;=\; z^{\prime}\;\frac{|{\bf W}|}{|{\bf W}_{\bot}|} \;=\; \frac{1}{2}\,\frac{k_{0}\mu\;{\cal R}^{2}(\kappa^{2} - \lambda\,k_{0}^{2})}{
 \sqrt{\kappa^{2} - \mu^{2}k_{0}^{2}}}, \]
 which leads to
 \begin{eqnarray} 
 \theta^{\prime}(s) & = & \frac{1}{2}\,k_{0}\mu\;\frac{(\kappa^{2} - \lambda k_{0}^{2})}{(\kappa^{2} - \mu^{2} k_{0}^{2})}  \nonumber \\
  & = & \frac{1}{2}\,k_{0}\mu\; \left[ 1 \;+\; \frac{k_{0}^{2}\,(\mu^{2} - \lambda)}{(\kappa^{2} - \mu^{2} k_{0}^{2})} \right].
 \label{eq:theta_prime} 
 \end{eqnarray}
The integral solution to this equation yields
 \begin{eqnarray}  
 \theta(s) & = & \mu\sqrt{q_{0}}\;\xi \nonumber \\
  &  & -\;i\,\mu\sqrt{q_{0}^{3}}\int_{\omega_{a}}^{i\xi + \omega_{a}}\frac{(\mu^{2} - \lambda)\,du}{\wp(u) - \wp(\Omega + \omega_{3})},
 \label{eq:theta_app}
 \end{eqnarray}
 where we used the Weierstrass solution \eqref{eq:kappa_PWJ}. The real-valued parameter $0 \leq \Omega(m) \leq \omega_{1}(m)$ is defined through the relation
 \begin{equation}
 \wp(\Omega + \omega_{3}) \;=\; q_{0}\left(\mu^{2} \;-\; \frac{2}{3}\,\lambda\right),
 \label{eq:wp_alpha}
 \end{equation}
 which yields the solution
\begin{equation} 
\Omega(m) \;=\; {\rm Re}\left\{\wp^{-1}\left[q_{0}\left(\mu^{2} \;-\; \frac{2}{3}\,\lambda\right) \right]\right\},
 \end{equation}
 where we used the fact that ${\rm Re}(\omega_{3}) \equiv 0$. We also find the useful identity
\begin{equation}
\wp^{\prime}(\Omega + \omega_{3}) \;\equiv\; -\,2\mu\sqrt{q_{0}^{3}}\;(\mu^{2} - \lambda),
\label{eq:p_prime_id}
\end{equation} 
which follows from $(\wp^{\prime})^{2} = 4\,\wp^{3} - g_{2}\,\wp - g_{3}$, with $\wp$ given by Eq.~\eqref{eq:wp_alpha} and $(g_{2}, g_{3})$ given by Eqs.~\eqref{eq:g2}-\eqref{eq:g3}. 

 \begin{figure}
\epsfysize=2in
\epsfbox{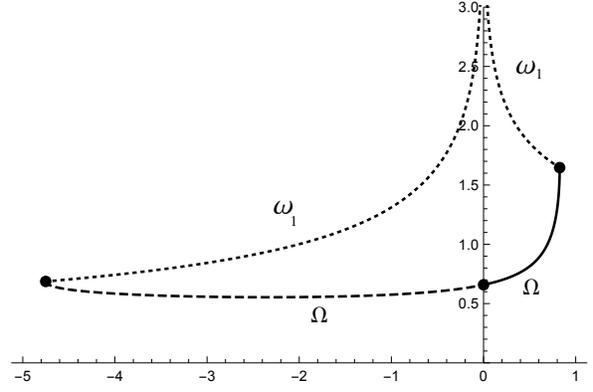}
\caption{Plot of $\Omega(m)$ as a function of $m$ in the classical range $0 \leq m \leq m_{0}^{-}$ (solid curve) and the extended range $m_{0}^{+} \leq m \leq 0$ (dashed curve). The real half-period $\omega_{1}(m)$ is also shown as a dotted curve, and $\Omega(m) = \omega_{1}(m)$ at $m = m_{0}^{-}$ and $m_{0}^{+}$.}
\label{fig:Omega_plot}
\end{figure}

When using these expressions, we therefore find
 \begin{equation}  
 \theta(\xi) \;=\; \mu\sqrt{q_{0}}\;\xi \;+\; \frac{i}{2}\int_{\omega_{a}}^{i\xi + \omega_{a}}\frac{\wp^{\prime}(v)\,du}{\wp(u) - \wp(v)},
 \label{eq:theta_int}
 \end{equation}
 where $v = \Omega + \omega_{3}$. In order to obtain an explicit solution from Eq.~\eqref{eq:theta_int}, we now use the identity \cite{Lawden}:
\begin{eqnarray*} 
\frac{\wp^{\prime}(v)}{\wp(u) - \wp(v)} & = & 2\,\zeta(v) \;+\; \zeta(u - v) \;-\; \zeta(u + v) \\
 & \equiv & 2\,\zeta(v) \;+\; \frac{d}{du}\left[\ln\left(\frac{\sigma(u - v)}{\sigma(u + v)}\right)\right],
 \end{eqnarray*}
 where the odd-parity Weierstrass zeta function $\zeta(u) = \sigma^{\prime}(u)/\sigma(u)$ is expressed in terms of the odd-parity Weierstrass sigma function $\sigma(u)$, so that 
we obtain the integral expression
  \[ \int\frac{\wp^{\prime}(v)\,du}{\wp(u) - \wp(v)} \;=\;  2u\;\zeta(v) \;+\; \ln\left( \frac{\sigma(u - v)}{\sigma(u+v)}\right). \]
  We note that the Weierstrass sigma function is not periodic but instead satisfies the relations
  \begin{equation}
  \left. \begin{array}{rcl}
  \sigma(u \pm 2\omega_{k}) & = & -\;\sigma(u)\;\exp\left[ \pm\,2\,\eta_{k}\frac{}{} (u \pm \omega_{k})\right] \\
  \sigma(u - \omega_{k}) & = & -\;\sigma(u + \omega_{k})\;\exp\left(-\,2\,\eta_{k}\frac{}{} u\right) 
  \end{array} \right\},
  \label{eq:sigma_per}
  \end{equation}
  where $\eta_{k} \equiv \zeta(\omega_{k})$.  Because $\sigma(u)$ vanishes at $u = 0$, Eq.~\eqref{eq:sigma_per} implies that it vanishes at the full periods: 
  $\sigma(2\omega_{k}) = 0$. 
  
  \begin{figure}
\epsfysize=1.8in
\epsfbox{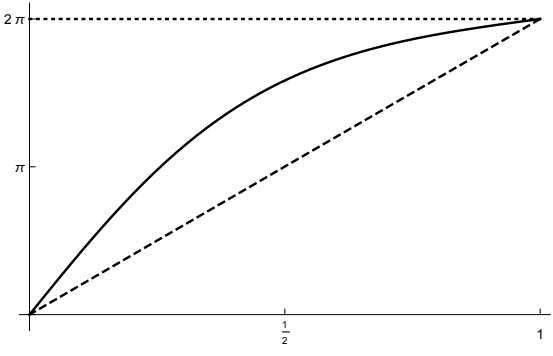}
\caption{Plot of the imaginary part of Eq.~\eqref{eq:Integral_net} as a function of $x = \Omega/\omega_{1}$ in the range $0 \leq x \leq 1$ for $m = 1/5$ (solid) and $m = 1$ (dashed).}
\label{fig:integral}
\end{figure}  

 The integral term in Eq.~\eqref{eq:theta_app} can thus be solved as
 \begin{eqnarray}
{\cal I}_{a}(\Omega,\xi) & \equiv & \int_{\omega_{a}}^{i\xi + \omega_{a}}\frac{\wp^{\prime}(v)\;du}{\wp(u) - \wp(v)} \label{eq:integral}  \\
 & = & 2\,i\,\xi\;\left[\zeta(v) + \zeta(\omega_{a}) \right] + \ln\left[ \frac{\sigma(\Omega - i\xi - \omega_{b})}{\sigma(\Omega + i\xi - \omega_{b})}\right],
\nonumber
 \end{eqnarray}
with $v \equiv \Omega + \omega_{3}$ and $\omega_{a} + \omega_{b} + \omega_{3} = 0$. We note that the function ${\cal I}_{a}(\Omega,\xi)$ takes values on the imaginary axis. In the classical range $\lambda < \lambda_{\Delta}$, we have $\omega_{a} = \omega_{1}$ and $\omega_{b} = \omega_{2}$, while in the extended range $\lambda > \lambda_{\Delta}$, we have $\omega_{a} = \omega_{2}$ and $\omega_{b} = \omega_{1}$. When Eq.~\eqref{eq:integral} is evaluated at $i\xi = 2\,\omega_{3}$, we find
\begin{eqnarray}
{\cal I}_{a}(\Omega, 2|\omega_{3}|) & = & 4\,\omega_{3}\;\zeta(\Omega + \omega_{3}) \;+\; 4\,\omega_{3}\;\zeta(\omega_{a}) \nonumber \\
 &  &+\; \ln\left[ \frac{\sigma(\Omega - 2\,\omega_{3} - \omega_{b})}{\sigma(\Omega + 2\,\omega_{3} - \omega_{b})}\right] \nonumber \\
  & = & {\cal I}(\Omega) \;+\; 4\;\left(\omega_{3}\;\eta_{a} \;-\frac{}{} \omega_{a}\;\eta_{3}\right),
\nonumber
 \end{eqnarray} 
 where we used $\omega_{3} + \omega_{b} = -\,\omega_{a}$ and we have defined
 \begin{equation}
{\cal I}(\Omega) \;\equiv\; 4\;\left[\omega_{3}\;\zeta(\Omega + \omega_{3}) \;-\frac{}{} (\Omega + \omega_{3})\;\zeta(\omega_{3}) \right].
\label{eq:Integral_net}
\end{equation}
In addition, we find $4(\omega_{3}\;\eta_{a} - \omega_{a}\;\eta_{3}) = +\,2i\,\pi$ ($a = 1$) or $-\,2i\,\pi$ ($a = 2$). Figure \ref{fig:integral} shows the imaginary part of ${\cal I}(\Omega)$ as a function of $x = \Omega/\omega_{1}$ in the range $0 \leq x \leq 1$.  When we substitute these results into Eq.~\eqref{eq:theta_app}, we obtain
 \begin{eqnarray}
 \theta(\xi) & = &  \mu\,\sqrt{q_{0}}\,\xi \;+\; \frac{i}{2}\;{\cal I}_{a}(\Omega,\xi) \nonumber \\
  & = & \xi\;\left[ \mu\,\sqrt{q_{0}} \;-\frac{}{} \zeta(\Omega + \omega_{3}) \;+\; \zeta(\omega_{3})\right]  \label{eq:theta_solution_m} \\
  &  &+\;  \frac{i}{2}\;\ln\left[\frac{\sigma(i\xi - \Omega - \omega_{b})\;\sigma(\Omega-\omega_{b})}{\sigma(-\Omega - \omega_{b})\;\sigma(i\xi + \Omega - \omega_{b})}\right],
\nonumber
\end{eqnarray}

 \begin{figure}
\epsfysize=1.5in
\epsfbox{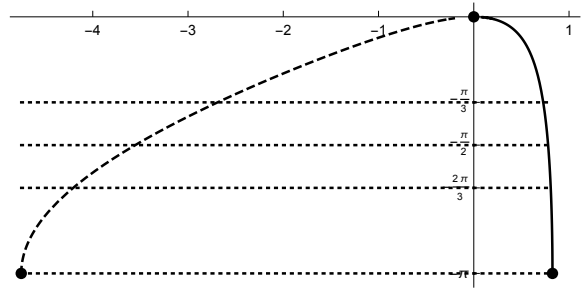}
\caption{Plot of $\Delta\theta(m)$ as a function of $m$ in the classical range $0 \leq m \leq m_{0}^{-}$ (solid curve) and the extended range $m_{0}^{+} \leq m \leq 0$ (dashed curve). Representative dotted lines at fractional values of $-\pi$ are shown at $-\pi/3$, $-\pi/2$, and $-2\pi/3$.}
\label{fig:Delta_plot}
\end{figure}

We can now define the azimuthal angular increment
\begin{eqnarray}
\Delta\theta(m) & \equiv & \theta(\xi + 2\,|\omega_{3}|) \;-\; \theta(\xi)  \label{eq:Delta_theta_m} \\
 & = & 2\,|\omega_{3}|\;\left[ \mu\,\sqrt{q_{0}} \;-\frac{}{} \zeta(\Omega + \omega_{3}) \;+\; \zeta(\omega_{3})\right] \nonumber \\
 &  + & \frac{i}{2}\;\ln\left[\frac{\sigma(i\xi - \Omega - \omega_{b} + 2\,\omega_{3})\;\sigma(i\xi + \Omega-\omega_{b})}{\sigma(i\xi -\Omega - \omega_{b})\;
\sigma(i\xi + \Omega - \omega_{b} + 2\,\omega_{3})}\right] \nonumber \\
 & = & 2\,\mu\sqrt{q_{0}}\;|\omega_{3}| \nonumber \\
  &  &+\; 2i\left[ \omega_{3}\;\zeta(\Omega + \omega_{3}) \;-\frac{}{} (\Omega + \omega_{3})\;\zeta(\omega_{3}) \right].
\nonumber
 \end{eqnarray}
 At $m = m_{0}^{\pm}$, we use the fact that $\mu\,q_{0}$ vanish and $\Omega = \omega_{1}$, to obtain the limiting angular increment
 \begin{eqnarray*} 
 \Delta\theta(m_{0}^{\pm}) & = & 2i\,\left[ \omega_{3}\;\zeta(\omega_{1} + \omega_{3}) \;-\frac{}{} (\omega_{1} + \omega_{3})\;\zeta(\omega_{3}) \right] \\
  & = & 2i\;\left[\omega_{2}\;\zeta(\omega_{3}) \;-\frac{}{} \omega_{3}\;\zeta(\omega_{2}) \right] \\
   & = & 2i\;\left(\frac{i\pi}{2}\right) \;=\; -\;\pi, 
   \end{eqnarray*}
 as can be seen in Fig.~\ref{fig:Delta_plot}, while
 \begin{eqnarray*} 
 \Delta\theta(0) & = & \frac{\pi}{\sqrt{3}} \;+\; 2i\;\left[ \frac{i\pi}{2}\;\zeta\left(\Omega_{0}^{+}\right) - \Omega_{0}^{+}\;\zeta\left(\frac{i\pi}{2}\right) \right] \\
& = & \frac{\pi}{\sqrt{3}} \;-\; \frac{\pi}{\sqrt{3}} \;=\; 0,
\end{eqnarray*}
where $\Omega_{0}^{+} \equiv \Omega_{0} + i\pi/2$, with $\Omega_{0} = {\rm arctanh}(1/\sqrt{3})$.
 
\begin{figure}
\epsfysize=2.5in
\epsfbox{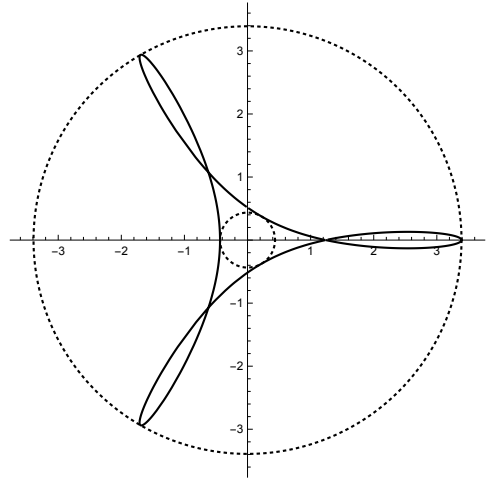}
\caption{Parametric plot of $y(\xi) = \rho(\xi)\,\sin\theta(\xi)$ versus $x(\xi) = \rho(\xi)\,\cos\theta(\xi)$ for the case $(\lambda,\nu) = (0.422531..., 0.0842782...)$ for which $\Delta\theta = -\,2\pi/3$.}
\label{fig:xy_parametric}
\end{figure}

We immediately conclude from Fig.~\ref{fig:Delta_plot}, where $\Delta\theta(m)$ is plotted in the range $m_{0}^{+} \leq m \leq m_{0}^{-}$, that periodic solutions in $(\rho,z,\theta)$ exist whenever $\Delta\theta(m)$ intersects lines at $-\,{\sf p}\pi/{\sf q}$, where $({\sf p}, {\sf q} \geq {\sf p})$ are integers. Hence, since $\theta(4\,{\sf q}{\sf K}/{\sf p}; m({\sf p}, {\sf q})) = -\,2\pi$, the cylindrical coordinates $(\rho,z)$ will return to their initial values only if $4\,{\sf q}{\sf K}/{\sf p} = 2\,\ell\,{\sf K}$, where $\ell$ is an integer. We therefore find ${\sf q} = \ell\,{\sf p}/2$ and, thus, 
\begin{equation}
\Delta\theta(m) \;=\; -\,{\sf p}\pi/{\sf q} = -\,2\pi/\ell(m). 
\label{eq:DeltaTheta_ell}
\end{equation}
The case ${\sf p} = 2$, ${\sf q} = 3$, and $\ell = 3$ is shown in Fig.~\ref{fig:xy_parametric}.

Lastly, we can also see from Fig.~\ref{fig:Delta_plot} that $\Delta\theta(m)$ satisfies the modulus symmetry
\begin{equation}
\Delta\theta[n(m)] \;=\; \Delta\theta(m),
\label{eq:Delta_theta_m}
\end{equation}
where $n(m) = -m/(1-m)$. 
 
\subsection{Cylindrical unit vectors}
  
  Once the azimuthal angle $\theta(s)$ is known, it is then possible to construct the rotating unit vectors
  \begin{equation}
  \left. \begin{array}{rcl}
  \wh{\rho}(s) & \equiv & \cos\theta(s)\;\wh{\sf x} \;+\; \sin\theta(s)\;\wh{\sf y} \\
   &  & \\
  \wh{\theta}(s) & \equiv & -\,\sin\theta(s)\;\wh{\sf x} \;+\; \cos\theta(s)\;\wh{\sf y}
  \end{array} \right\},
  \end{equation}
  where the fixed unit vectors $(\wh{\sf x}, \wh{\sf y})$ define a plane perpendicular to the unit vector $\wh{\sf z}$.
  
  \subsubsection{Frenet-Serret frame}
  
  Our task is now to write expressions for the Frenet-Serret unit vectors $(\wh{\sf t}, \wh{\sf n}, \wh{\sf b})$ in terms of the cylindrical unit vectors $(\wh{\rho},\wh{\theta},\wh{\sf z})$. First, using Eq.~\eqref{eq:z_def}, Eqs.~\eqref{eq:theta_def}-\eqref{eq:rho_def} become
 \begin{eqnarray} 
  \wh{\theta} & = & \frac{\wh{\sf z} - \gamma^{-1}\wh{\sf b}}{\sqrt{1 - 1/\gamma^{2}}} \;=\; \frac{\gamma\,(\alpha\,\wh{\sf t} + \beta\,\wh{\sf n}) - (1 - \gamma^{2})\,\wh{\sf b}}{\sqrt{1 - \gamma^{2}}},  \label{eq:theta_FS} \\
  \wh{\rho} & = & \frac{(\wh{\sf z}\btimes\wh{\sf b})}{\sqrt{1 - \gamma^{2}}} \;=\; \frac{(\beta\,\wh{\sf t} - \alpha\,\wh{\sf n})}{\sqrt{1 - \gamma^{2}}}. \label{eq:rho_FS}
 \end{eqnarray}
 Next, using Eq.~\eqref{eq:theta_FS}, with $\wh{\sf x}\bdot\wh{\theta} = -\,\sin\theta$ and $\wh{\sf y}\bdot\wh{\theta} = \cos\theta$, we now find
 \[ \left( \begin{array}{c}
 b_{x} \\
 b_{y}
 \end{array} \right) \;=\; \sqrt{1 - \gamma^{2}}\; \left( \begin{array}{c}
\sin\theta \\
-\,\cos\theta
 \end{array} \right), \]
 and, thus, the binormal unit vector is
 \begin{equation}
 \wh{\sf b} \;=\; \gamma\,\wh{\sf z} \;-\; \sqrt{1 - \gamma^{2}}\;\wh{\theta}.
 \label{eq:b_cyl}
 \end{equation}
 Using Eq.~\eqref{eq:rho_FS}, with $\wh{\sf x}\bdot\wh{\rho} = \cos\theta$ and $\wh{\sf y}\bdot\wh{\rho} = \sin\theta$, we also find
 \[ \left( \begin{array}{c}
\cos\theta \\
\sin\theta
 \end{array} \right) \;=\; \frac{1}{\sqrt{1 - \gamma^{2}}}\; \left( \begin{array}{c}
\beta\,t_{x} \;-\; \alpha\,n_{x} \\
\beta\,t_{y} \;-\; \alpha\,n_{y}
 \end{array} \right). \]
 If we now introduce the decompositions 
 \begin{eqnarray*}
 (t_{x},n_{x}) & = & (t_{\rho},n_{\rho})\,\cos\theta - (t_{\theta},n_{\theta})\,\sin\theta, \\
 (t_{y},n_{y}) & = & (t_{\rho},n_{\rho})\,\sin\theta + (t_{\theta},n_{\theta})\,\cos\theta,
 \end{eqnarray*}
 we readily find the relations
 \begin{equation}
 \left. \begin{array}{rcl}
 \beta\,t_{\rho} \;-\; \alpha\;n_{\rho} & = & \sqrt{1 - \gamma^{2}} \\
  &  & \\
  \beta\,t_{\theta} \;-\; \alpha\,n_{\theta} & = & 0
  \end{array} \right\}.
  \label{eq:tn_rel}
  \end{equation}
 Lastly, if we use Eq.~\eqref{eq:b_cyl}, with $\wh{\sf n} = \beta\,\wh{\sf z} + n_{\rho}\,\wh{\rho} + n_{\theta}\,\wh{\theta}$, we find
 \begin{eqnarray*}
 \wh{\sf t} \;=\; \wh{\sf n}\btimes\wh{\sf b} & = & -\;n_{\rho}\;\sqrt{1 - \gamma^{2}} \;\wh{\sf z} \;-\; \alpha\,n_{\rho}\;\wh{\theta} \\
  &  &+\; \left( \gamma\,n_{\theta} \;+\frac{}{} \beta\, \sqrt{1 - \gamma^{2}} \right)\wh{\rho},
  \end{eqnarray*} 
  and, hence, we obtain $\alpha = -\;n_{\rho}\;\sqrt{1 - \gamma^{2}}$, $t_{\theta} = -\; \alpha\,n_{\rho}$, and $t_{\rho} = \gamma\,n_{\theta} + \beta\, \sqrt{1 - \gamma^{2}}$.  By using the relations \eqref{eq:tn_rel}, we now easily obtain
  \begin{eqnarray}
  \wh{\sf t} & = & \alpha\,\wh{\sf z} \;+\; \frac{(\beta\,\wh{\rho} + \alpha\,\gamma\,\wh{\theta})}{\sqrt{1 - \gamma^{2}}}, \label{eq:t_cyl} \\
  \wh{\sf n} & = & \beta\,\wh{\sf z} \;+\; \frac{(-\,\alpha\,\wh{\rho} + \beta\,\gamma\,\wh{\theta})}{\sqrt{1 - \gamma^{2}}}. \label{eq:n_cyl}
  \end{eqnarray}
  Hence, the Frenet-Serret triad $(\wh{\sf t}, \wh{\sf n}, \wh{\sf b})$ is now completely expressed in terms of cylindrical geometry. In the limit of zero torsion $(\tau_{0} = 0 = \gamma)$, to be discussed in the next subsection, the binormal unit vector $\wh{\sf b} \equiv -\,\wh{\theta}$ is constant, and $\wh{\sf t} = \alpha\,\wh{\sf z} + \beta\,\wh{\rho}$ and $\wh{\sf n} = \beta\,\wh{\sf z} - \alpha\,\wh{\rho}$ are on the $(\rho,z)$-plane.
  
  \subsubsection{Darboux frame}
  
The Frenet-Serret triad is not unique along the curve ${\bf r}(s)$. The Darboux triad $(\wh{\sf T}, \wh{\sf N}, \wh{\sf B})$ is constructed from the Frenet-Serret triad $(\wh{\sf t}, \wh{\sf n}, \wh{\sf b})$ by a rotation through an angle $\Theta(s)$ about the $\wh{\sf t}$-axis:
  \begin{equation}
 \left. \begin{array}{rcl}
 \wh{\sf T} & = & \wh{\sf t} \\
 \wh{\sf N} & = & \cos\Theta\;\wh{\sf n} \;+\; \sin\Theta\;\wh{\sf b} \\
 \wh{\sf B} & = & -\,\sin\Theta\;\wh{\sf n} \;+\; \cos\Theta\;\wh{\sf b}
 \end{array} \right\},
 \label{eq:Darboux}
 \end{equation}
 from which we obtain the Darboux equations
 \begin{equation}
\left. \begin{array}{rcl}
 d\wh{\sf T}/ds & = & \kappa\,\cos\Theta\;\wh{\sf N} \;-\; \kappa\,\sin\Theta\;\wh{\sf B} \\
 d\wh{\sf N}/ds & = & -\,\kappa\,\cos\Theta\;\wh{\sf T} \;+\; (\tau + d\Theta/ds)\;\wh{\sf B} \\
 d\wh{\sf B}/ds & = & \kappa\,\sin\Theta\;\wh{\sf T} \;-\; (\tau + d\Theta/ds)\;\wh{\sf N}
 \end{array} \right\}.
 \label{eq:D_eq}
 \end{equation}
 Here, the Darboux angle $\Theta(s)$ measures the deviation of the elastica curve from a geodesic curve on the toroidal surface on which it lies.
 
 If we substitute the Frenet-Serret formulas \eqref{eq:b_cyl} and \eqref{eq:n_cyl} into the expression for $\wh{\sf N}$, we find the $\wh{\sf t}$-component:
 \[ \wh{\sf N}\bdot\wh{\sf t} \;=\; \alpha\;\left(1 - \sqrt{1 - \gamma^{2}}\right) \left(\beta\;\cos\Theta \;+\frac{}{} \gamma\;\sin\Theta\right), \]
 which is required to vanish because $\wh{\sf N}\bdot\wh{\sf t} = \wh{\sf N}\bdot\wh{\sf T} \equiv 0$. Hence, the Darboux angle $\Theta(s)$ is defined as
 \begin{eqnarray}
 \Theta(s) & \equiv & {\rm arctan}\left(-\;\frac{\beta(s)}{\gamma(s)}\right) \;=\; {\rm arctan}\left( -\;\frac{(\kappa^{2})^{\prime}}{k_{0}^{3}\nu}\right) \nonumber \\
  & = & {\rm arctan}\left[ -\;\frac{i}{2\nu}\;\wp^{\prime}\left(i\;\frac{k_{0}s}{2} + \omega_{a} \right) \right],
 \label{eq:Theta_D}
 \end{eqnarray}
where $\Theta(0) = 0$. Hence, we may express the Weierstrass derivative
\begin{equation}
\wp^{\prime}\left(i\;\frac{k_{0}s}{2} + \omega_{a}\right) \;\equiv\; 2i\,\nu\;\tan\Theta(s)
\label{eq:wp_prime_Darboux}
\end{equation}
in terms of the Darboux angle $\Theta(s)$, so that $(\kappa^{2})^{\prime} = -\,k_{0}^{3}\nu\;\tan\Theta(s)$.

According to Eq.~\eqref{eq:D_eq}, we obtain an expression for the geodesic curvature 
 \[ \kappa_{g}(s) \;\equiv\; \kappa(s)\,\cos\Theta(s) \;=\; \frac{\kappa(s)}{\sqrt{1 \;+\; [(\kappa^{2})^{\prime}]^{2}/(k_{0}^{6}\nu^{2})}}, \]
 which measures curvature in relation to geodesics, where
 \begin{eqnarray*} 
 \left[\left(\kappa^{2}\right)^{\prime}\right]^{2} + k_{0}^{6}\nu^{2} & =  & \kappa^{2}\;k_{0}^{4}\,\left( (1 - \lambda)^{2} +\frac{}{} \nu^{2} \right) \\
  &  &-\; \kappa^{2} \left(\kappa^{2} \;-\frac{}{} \lambda\, k_{0}^{2}\right)^{2}, 
  \end{eqnarray*}
 so that
 \begin{equation}
\kappa_{g}(s) = k_{0}^{3}\nu \left[ k_{0}^{4} \left( (1 - \lambda)^{2} \;+\frac{}{} \nu^{2}\right) - \left( \kappa^{2} \;-\frac{}{} \lambda\,k_{0}^{2}\right)^{2} \right]^{-\,\frac{1}{2}}.
 \end{equation}
 We also find expressions for the normal curvature $\kappa_{n}(s) \equiv -\,\kappa(s)\,\sin\Theta(s)$ and the geodesic (relative) torsion is $\tau_{r}(s) \equiv \tau(s) + d\Theta(s)/ds$.
  
 \section{\label{sec:summary}Summary}
 
 In summary, we have shown that the Jacobi elliptic solution \eqref{eq:kappa_Jacobi} for the Frenet-Serret curvature $\kappa(s)$ can be used to construct equivalent elastica knots associated with constant curvature and torsion functionals. Hence, for a normalized curvature functional \eqref{eq:F_S}  evaluated as $\ov{\cal F} = 2 < \pi$, for example, we find two equivalent Jacobi elliptic solutions with moduli $m_{-} = 0.751...$ and $m_{+} = -\,3.02...$, where $n(m_{+}) = -m_{+}/(1-m_{+}) = m_{-}$. These equivalent Jacobi elliptic solutions also have the same numerical values for the normalized averaged torsion $\langle\tau\rangle = 0.601...$ and total torsion $T = 0.288...$.
 
\appendix

\section{\label{sec:W}Weierstrass Parametrization of the Curvature Solution}
 
 The Weierstrass and Jacobi elliptic solutions to the curvature equation presented in Sec.~\ref{sec:LS_J} were parametrized in terms of the Langer-Singer parameters $(m,q_{0})$. In this Appendix, we introduce a new parametrization based on the curvature parameters $(\lambda,\nu^{2})$ in the half plane $-\infty < \lambda < \infty$ and $\nu^{2} \geq 0$. Using our new parametrization, we show how the scale parameter $q_{0}$ can be completely eliminated from the parametric representation of elastica knots.
 
 \subsection{Weierstrass parametrization}
 
 First, we pointed out that the invariant functions \eqref{eq:g2}-\eqref{eq:Delta_general} were homogeneous functions of the scale parameter $q_{0}$. Next, we note that the Weierstrass elliptic function $\wp(u; g_{2}, g_{2})$ is invariant under the homogeneity transformation \cite{NIST_Weierstrass}
 \begin{equation} 
 t^{2}\;\wp\left( t\,u; t^{-4}\,g_{2}, t^{-6}\,g_{3}\right) \;=\; \wp(u; g_{2}\,g_{3}),
 \label{eq:hom}
 \end{equation} 
 where $t$ is an arbitrary number (real or complex). Thus, if we use $t = q_{0}^{-\frac{1}{2}}$, with $\xi = \ov{\xi}\,q_{0}^{-\frac{1}{2}}$, $g_{2} = \ov{g}_{2}(\lambda,\nu)\,q_{0}^{2}$, and $g_{3} =  \ov{g}_{3}(\lambda,\nu)\,q_{0}^{3}$ [where $(\ov{g}_{2}, \ov{g}_{3})$ are given in Eqs.~\eqref{eq:g2}-\eqref{eq:g3}], then the transformation \eqref{eq:hom} yields
 \begin{equation} 
 q_{0}^{-1}\wp\left(q_{0}^{-\frac{1}{2}}\,\ov{u}; q_{0}^{2}\,\ov{g}_{2}, q_{0}^{3}\,\ov{g}_{3}\right) \;=\; \wp(\ov{u}; \ov{g}_{2}, \ov{g}_{3}),
 \label{eq:hom_q0} 
 \end{equation}
 with  $\ov{u} = i\,k_{0}s/2 + \ov{\varphi}_{0}$ and $\ov{\varphi}_{0} = \varphi_{0}\,\sqrt{q_{0}}$, while the new cubic roots are defined from Eqs.~\eqref{eq:e1_chi}-\eqref{eq:e23_chi} as  $\ov{\sf e}_{k}(\lambda,\nu) = {\sf e}_{k}/q_{0}$. We have, therefore, eliminated the parameter $q_{0}$ and we are left with the two independent parameters $\lambda$ and $\nu$. The half-periods $\ov{\omega}_{k}(\lambda,\nu) \equiv \sqrt{q_{0}}\,\omega_{k}$, defined by the relations $\wp(\ov{\omega}_{k}; \ov{g}_{2}, \ov{g}_{3}) = \ov{\sf e}_{k}$, also depend on both parameters.
 
The two-parameter Weierstrass solution $\kappa^{2}(s; \lambda,\nu)$ is obtained from Eq.~\eqref{eq:kappa_PWJ} as
 \begin{eqnarray}
\kappa^{2}(s) & = & k_{0}^{2} \left[ \frac{2}{3}\,\lambda + \wp\left( i\,\ov{\xi} + \ov{\omega}_{a};\; \ov{g}_{2}(\lambda,\nu), \ov{g}_{3}(\lambda,\nu)\right) \right] 
\nonumber \\
 & = & k_{0}^{2} \left[ 1 \;+\; \wp\left( i\,\ov{\xi} + \ov{\omega}_{a};\; \ov{g}_{2}, \ov{g}_{3}\right) \;-\; \ov{\sf e}_{a} \right],
 \label{eq:kappa_PW}
 \end{eqnarray}
where $\ov{\xi} = k_{0}s/2$ and the scale parameter $q_{0}$ has now completely disappeared from our Weierstrass solution. 

\subsection{Weierstrass cubic roots and half-periods}

According to Fig.~\ref{fig:ek_W}, in the classical range $\lambda < \lambda_{\Delta}$, the ordered cubic roots $\ov{\sf e}_{3}^{-} < \ov{\sf e}_{2}^{-} \leq \ov{\sf e}_{1}^{-}$ are
\begin{equation}
\left. \begin{array}{rcl}
\ov{\sf e}_{1}^{-}(\lambda,\nu) & = & 1 \;-\; \frac{2}{3}\,\lambda \\
\ov{\sf e}_{2}^{-}(\lambda,\nu) & = & \left(\frac{1}{3}\,\lambda \;-\; \frac{1}{2}\right) \;+\; \frac{1}{2}\,\delta(\lambda,\nu) \\
 &  & \\
\ov{\sf e}_{3}^{-}(\lambda,\nu) & = & \left(\frac{1}{3}\,\lambda \;-\; \frac{1}{2}\right) \;-\; \frac{1}{2}\,\delta(\lambda,\nu)
\end{array} \right\},
\label{eq:roots_I}
\end{equation}
so that the two-parameter Weierstrass solution \eqref{eq:kappa_PW} becomes
 \begin{equation}
\kappa_{-}^{2}(s) \;=\; k_{0}^{2} \left\{ 1 \;+\frac{}{} \left[ \wp\left(  i\,\ov{\xi} + \ov{\omega}_{1}^{-};\; \ov{g}_{2}, \ov{g}_{3}\right) \;-\frac{}{}
\ov{\sf e}_{1}^{-} \right] \right\}.
 \label{eq:kappa_PW_I}
 \end{equation}
The Jacobi modulus is
\begin{equation}
\ov{p}^{-}(\lambda,\nu) \;=\; \frac{\delta(\lambda,\nu)}{(3/2 - \lambda) \;+\; \frac{1}{2}\,\delta(\lambda,\nu)},
\end{equation}
and the half-periods $\ov{\omega}_{1}^{-}$ and $\ov{\omega}_{3}^{-}$ are
\begin{equation}
\left( \ov{\omega}_{1}^{-},\frac{}{}\ov{\omega}_{3}^{-}\right) \;=\; \frac{1}{\sqrt{\ov{\sf e}_{1}^{-} - \ov{\sf e}_{3}^{-}}} \left( {\sf K}(\ov{p}^{-}),\frac{}{} i\,{\sf K}(1 - \ov{p}^{-})\right),
\end{equation}
where $\ov{\sf e}_{1}^{-} - \ov{\sf e}_{3}^{-} = (\frac{3}{2} - \lambda) + \frac{1}{2}\,\delta(\lambda,\nu)$. The two-parameter Jacobi elliptic solution, on the other hand, is
\begin{equation}
\kappa_{-}^{2}(s) \;=\; k_{0}^{2}  \left[ 1 \;-\; (\ov{\sf e}_{1}^{-} - \ov{\sf e}_{2}^{-})\;{\rm sn}^{2}\left( \ov{\xi}^{-}|\,1 - p^{-} \right) \right],
\end{equation}
where $\ov{\xi}^{-} \equiv \ov{\xi}\,\sqrt{\ov{\sf e}_{1}^{-} - \ov{\sf e}_{3}^{-}}$ and $\ov{\sf e}_{1}^{-} - \ov{\sf e}_{2}^{-} = (\frac{3}{2} - \lambda) - \frac{1}{2}\,\delta(\lambda,\nu)$.
 
In the extended range $\lambda > \lambda_{\Delta}$, on the other hand, the ordered cubic roots $\ov{\sf e}_{3}^{+} < \ov{\sf e}_{2}^{+} < \ov{\sf e}_{1}^{+}$ are
\begin{equation}
\left. \begin{array}{rcl}
\ov{\sf e}_{1}^{+}(\lambda,\nu) & = & \left( \frac{1}{3}\,\lambda \;-\; \frac{1}{2}\right) \;+\; \frac{1}{2}\,\delta(\lambda,\nu) \\
\ov{\sf e}_{2}^{+}(\lambda,\nu) & = & 1 \;-\; \frac{2}{3}\,\lambda \\
\ov{\sf e}_{3}^{+}(\lambda,\nu) & = & \left( \frac{1}{3}\,\lambda \;-\; \frac{1}{2}\right) \;-\; \frac{1}{2}\,\delta(\lambda,\nu)
\end{array} \right\},
\label{eq:roots_II}
\end{equation} 
so that the two-parameter Weierstrass solution \eqref{eq:kappa_PW} becomes
 \begin{equation}
\kappa_{+}^{2}(s) \;=\; k_{0}^{2} \left\{ 1 \;+\frac{}{} \left[ \wp\left(  i\,\ov{\xi} + \ov{\omega}_{2}^{+};\; \ov{g}_{2}, \ov{g}_{3}\right) \;-\; 
\ov{\sf e}_{2}^{+}\right] \right\}.
 \label{eq:kappa_PW_II}
 \end{equation}
 The Jacobi modulus is
\begin{equation}
\ov{p}^{+}(\lambda,\nu) \;=\; \frac{(3/2 - \lambda)}{\delta(\lambda,\nu)} \;+\; \frac{1}{2},
\label{eq:ov_p_plus}
\end{equation}
and the half-periods $\ov{\omega}_{1}^{+}$ and $\ov{\omega}_{3}^{+}$ are
\begin{equation}
\left( \ov{\omega}_{1}^{+},\frac{}{}\ov{\omega}_{3}^{+}\right) \;=\; \frac{1}{\sqrt{\ov{\sf e}_{1}^{+} - \ov{\sf e}_{3}^{+}}} \left( {\sf K}(\ov{p}^{+}),\frac{}{} i\,{\sf K}(1 - \ov{p}^{+})\right),
\end{equation}
where $\ov{\sf e}_{1}^{+} - \ov{\sf e}_{3}^{+} = \delta(\lambda,\nu)$. The two-parameter Jacobi elliptic solution, on the other hand, is
\begin{equation}
\kappa_{+}^{2}(s) \;=\; k_{0}^{2}  \left[ 1 + \ov{p}^{+}(\ov{\sf e}_{1}^{+} - \ov{\sf e}_{2}^{+}){\rm sd}^{2}(\ov{\xi}^{+}\,|\,1 - p^{+}) \right],
\end{equation}
where $\ov{\xi}^{+} \equiv \ov{\xi}\,\sqrt{\ov{\sf e}_{1}^{+} - \ov{\sf e}_{3}^{+}}$ and $\ov{\sf e}_{1}^{+} - \ov{\sf e}_{2}^{+} = (\lambda - \frac{3}{2}) + \frac{1}{2}\,\delta(\lambda,\nu)$. We note that, in the limit $\nu \gg 1$, the Jacobi modulus \eqref{eq:ov_p_plus} becomes $\ov{p}^{+} \rightarrow \frac{1}{2}$ for all values of $\lambda$ (see Fig.~\ref{fig:m_W}).

\end{document}